\documentclass[aps,prl,10pt,twocolumn,superscriptaddress,groupedaddress,amsmath,amssymb,nofootinbib,preprintnumbers]{revtex4-2} 

\usepackage{graphicx}
\usepackage{dcolumn}
\usepackage{bm}
\usepackage{multirow}%
\usepackage{xcolor}
\usepackage[colorlinks=true,linkcolor=blue,urlcolor=blue,citecolor = black]{hyperref}
\usepackage[mathlines]{lineno}
\usepackage{enumitem} 
\hypersetup{
    colorlinks=true,
    linkcolor=cyan,
    filecolor=magenta,      
    urlcolor=cyan,
    citecolor=violet,
}
\usepackage{mathptmx}
\usepackage{cleveref}
\makeatletter
\g@addto@macro\bfseries{\boldmath}
\makeatother

\newcommand{\Offline}{$\overline{\textrm{Off}}$\hspace{.05em}\protect\raisebox{.4ex}{$\protect\underline{\textrm{line}}\mbox{ }$}}

\begin{document}

\title{A search for the anomalous events detected by ANITA\\ using the Pierre Auger Observatory}

\email{spokespersons@auger.org}



\author{
A.~Abdul Halim$^{13}$,
P.~Abreu$^{73}$,
M.~Aglietta$^{55,53}$,
I.~Allekotte$^{1}$,
K.~Almeida Cheminant$^{71}$,
A.~Almela$^{7,12}$,
R.~Aloisio$^{46,47}$,
J.~Alvarez-Mu\~niz$^{79}$,
J.~Ammerman Yebra$^{79}$,
G.A.~Anastasi$^{59,48}$
% ...
}
\affiliation{}
\collaboration{Pierre Auger Collaboration}
\author{R.~Prechelt$^{\,o}$}
\author{A.~Romero-Wolf$^{\,p}$}
\author{S.~Wissel$^{\,l}$}
\author{A.~Zeolla$^{\,l}$}
\vspace*{5mm}
\affiliation{
\begin{description}[labelsep=0.2em,align=right,labelwidth=0.7em,labelindent=0em,leftmargin=2em,noitemsep,before={\renewcommand\makelabel[1]{##1 }}]
\item[$^{1}$] Centro At\'omico Bariloche and Instituto Balseiro (CNEA-UNCuyo-CONICET), San Carlos de Bariloche, Argentina
\item[$^{2}$] Departamento de F\'\i{}sica and Departamento de Ciencias de la Atm\'osfera y los Oc\'eanos, FCEyN, Universidad de Buenos Aires and CONICET, Buenos Aires, Argentina
\item[$^{3}$] IFLP, Universidad Nacional de La Plata and CONICET, La Plata, Argentina
\item[$^{4}$] Instituto de Astronom\'\i{}a y F\'\i{}sica del Espacio (IAFE, CONICET-UBA), Buenos Aires, Argentina
\item[$^{5}$] Instituto de F\'\i{}sica de Rosario (IFIR) -- CONICET/U.N.R.\ and Facultad de Ciencias Bioqu\'\i{}micas y Farmac\'euticas U.N.R., Rosario, Argentina
\item[$^{6}$] Instituto de Tecnolog\'\i{}as en Detecci\'on y Astropart\'\i{}culas (CNEA, CONICET, UNSAM), and Universidad Tecnol\'ogica Nacional -- Facultad Regional Mendoza (CONICET/CNEA), Mendoza, Argentina
\item[$^{7}$] Instituto de Tecnolog\'\i{}as en Detecci\'on y Astropart\'\i{}culas (CNEA, CONICET, UNSAM), Buenos Aires, Argentina
\item[$^{8}$] International Center of Advanced Studies and Instituto de Ciencias F\'\i{}sicas, ECyT-UNSAM and CONICET, Campus Miguelete -- San Mart\'\i{}n, Buenos Aires, Argentina
\item[$^{9}$] Laboratorio Atm\'osfera -- Departamento de Investigaciones en L\'aseres y sus Aplicaciones -- UNIDEF (CITEDEF-CONICET), Argentina
\item[$^{10}$] Observatorio Pierre Auger, Malarg\"ue, Argentina
\item[$^{11}$] Observatorio Pierre Auger and Comisi\'on Nacional de Energ\'\i{}a At\'omica, Malarg\"ue, Argentina
\item[$^{12}$] Universidad Tecnol\'ogica Nacional -- Facultad Regional Buenos Aires, Buenos Aires, Argentina
\item[$^{13}$] University of Adelaide, Adelaide, S.A., Australia
\item[$^{14}$] Universit\'e Libre de Bruxelles (ULB), Brussels, Belgium
\item[$^{15}$] Vrije Universiteit Brussels, Brussels, Belgium
\item[$^{16}$] Centro Brasileiro de Pesquisas Fisicas, Rio de Janeiro, RJ, Brazil
\item[$^{17}$] Centro Federal de Educa\c{c}\~ao Tecnol\'ogica Celso Suckow da Fonseca, Petropolis, Brazil
\item[$^{18}$] Instituto Federal de Educa\c{c}\~ao, Ci\^encia e Tecnologia do Rio de Janeiro (IFRJ), Brazil
\item[$^{19}$] Universidade de S\~ao Paulo, Escola de Engenharia de Lorena, Lorena, SP, Brazil
\item[$^{20}$] Universidade de S\~ao Paulo, Instituto de F\'\i{}sica de S\~ao Carlos, S\~ao Carlos, SP, Brazil
\item[$^{21}$] Universidade de S\~ao Paulo, Instituto de F\'\i{}sica, S\~ao Paulo, SP, Brazil
\item[$^{22}$] Universidade Estadual de Campinas (UNICAMP), IFGW, Campinas, SP, Brazil
\item[$^{23}$] Universidade Estadual de Feira de Santana, Feira de Santana, Brazil
\item[$^{24}$] Universidade Federal de Campina Grande, Centro de Ciencias e Tecnologia, Campina Grande, Brazil
\item[$^{25}$] Universidade Federal do ABC, Santo Andr\'e, SP, Brazil
\item[$^{26}$] Universidade Federal do Paran\'a, Setor Palotina, Palotina, Brazil
\item[$^{27}$] Universidade Federal do Rio de Janeiro, Instituto de F\'\i{}sica, Rio de Janeiro, RJ, Brazil
\item[$^{28}$] Universidade Federal do Rio de Janeiro (UFRJ), Observat\'orio do Valongo, Rio de Janeiro, RJ, Brazil
\item[$^{29}$] Universidade Federal Fluminense, EEIMVR, Volta Redonda, RJ, Brazil
\item[$^{30}$] Universidad de Medell\'\i{}n, Medell\'\i{}n, Colombia
\item[$^{31}$] Universidad Industrial de Santander, Bucaramanga, Colombia
\item[$^{32}$] Charles University, Faculty of Mathematics and Physics, Institute of Particle and Nuclear Physics, Prague, Czech Republic
\item[$^{33}$] Institute of Physics of the Czech Academy of Sciences, Prague, Czech Republic
\item[$^{34}$] Palacky University, Olomouc, Czech Republic
\item[$^{35}$] CNRS/IN2P3, IJCLab, Universit\'e Paris-Saclay, Orsay, France
\item[$^{36}$] Laboratoire de Physique Nucl\'eaire et de Hautes Energies (LPNHE), Sorbonne Universit\'e, Universit\'e de Paris, CNRS-IN2P3, Paris, France
\item[$^{37}$] Univ.\ Grenoble Alpes, CNRS, Grenoble Institute of Engineering Univ.\ Grenoble Alpes, LPSC-IN2P3, 38000 Grenoble, France
\item[$^{38}$] Universit\'e Paris-Saclay, CNRS/IN2P3, IJCLab, Orsay, France
\item[$^{39}$] Bergische Universit\"at Wuppertal, Department of Physics, Wuppertal, Germany
\item[$^{40}$] Karlsruhe Institute of Technology (KIT), Institute for Experimental Particle Physics, Karlsruhe, Germany
\item[$^{41}$] Karlsruhe Institute of Technology (KIT), Institut f\"ur Prozessdatenverarbeitung und Elektronik, Karlsruhe, Germany
\item[$^{42}$] Karlsruhe Institute of Technology (KIT), Institute for Astroparticle Physics, Karlsruhe, Germany
\item[$^{43}$] RWTH Aachen University, III.\ Physikalisches Institut A, Aachen, Germany
\item[$^{44}$] Universit\"at Hamburg, II.\ Institut f\"ur Theoretische Physik, Hamburg, Germany
\item[$^{45}$] Universit\"at Siegen, Department Physik -- Experimentelle Teilchenphysik, Siegen, Germany
\item[$^{46}$] Gran Sasso Science Institute, L'Aquila, Italy
\item[$^{47}$] INFN Laboratori Nazionali del Gran Sasso, Assergi (L'Aquila), Italy
\item[$^{48}$] INFN, Sezione di Catania, Catania, Italy
\item[$^{49}$] INFN, Sezione di Lecce, Lecce, Italy
\item[$^{50}$] INFN, Sezione di Milano, Milano, Italy
\item[$^{51}$] INFN, Sezione di Napoli, Napoli, Italy
\item[$^{52}$] INFN, Sezione di Roma ``Tor Vergata'', Roma, Italy
\item[$^{53}$] INFN, Sezione di Torino, Torino, Italy
\item[$^{54}$] Istituto di Astrofisica Spaziale e Fisica Cosmica di Palermo (INAF), Palermo, Italy
\item[$^{55}$] Osservatorio Astrofisico di Torino (INAF), Torino, Italy
\item[$^{56}$] Politecnico di Milano, Dipartimento di Scienze e Tecnologie Aerospaziali , Milano, Italy
\item[$^{57}$] Universit\`a del Salento, Dipartimento di Matematica e Fisica ``E.\ De Giorgi'', Lecce, Italy
\item[$^{58}$] Universit\`a dell'Aquila, Dipartimento di Scienze Fisiche e Chimiche, L'Aquila, Italy
\item[$^{59}$] Universit\`a di Catania, Dipartimento di Fisica e Astronomia ``Ettore Majorana``, Catania, Italy
\item[$^{60}$] Universit\`a di Milano, Dipartimento di Fisica, Milano, Italy
\item[$^{61}$] Universit\`a di Napoli ``Federico II'', Dipartimento di Fisica ``Ettore Pancini'', Napoli, Italy
\item[$^{62}$] Universit\`a di Palermo, Dipartimento di Fisica e Chimica ''E.\ Segr\`e'', Palermo, Italy
\item[$^{63}$] Universit\`a di Roma ``Tor Vergata'', Dipartimento di Fisica, Roma, Italy
\item[$^{64}$] Universit\`a Torino, Dipartimento di Fisica, Torino, Italy
\item[$^{65}$] Benem\'erita Universidad Aut\'onoma de Puebla, Puebla, M\'exico
\item[$^{66}$] Unidad Profesional Interdisciplinaria en Ingenier\'\i{}a y Tecnolog\'\i{}as Avanzadas del Instituto Polit\'ecnico Nacional (UPIITA-IPN), M\'exico, D.F., M\'exico
\item[$^{67}$] Universidad Aut\'onoma de Chiapas, Tuxtla Guti\'errez, Chiapas, M\'exico
\item[$^{68}$] Universidad Michoacana de San Nicol\'as de Hidalgo, Morelia, Michoac\'an, M\'exico
\item[$^{69}$] Universidad Nacional Aut\'onoma de M\'exico, M\'exico, D.F., M\'exico
\item[$^{70}$] Universidad Nacional de San Agustin de Arequipa, Facultad de Ciencias Naturales y Formales, Arequipa, Peru
\item[$^{71}$] Institute of Nuclear Physics PAN, Krakow, Poland
\item[$^{72}$] University of \L{}\'od\'z, Faculty of High-Energy Astrophysics,\L{}\'od\'z, Poland
\item[$^{73}$] Laborat\'orio de Instrumenta\c{c}\~ao e F\'\i{}sica Experimental de Part\'\i{}culas -- LIP and Instituto Superior T\'ecnico -- IST, Universidade de Lisboa -- UL, Lisboa, Portugal
\item[$^{74}$] ``Horia Hulubei'' National Institute for Physics and Nuclear Engineering, Bucharest-Magurele, Romania
\item[$^{75}$] Institute of Space Science, Bucharest-Magurele, Romania
\item[$^{76}$] Center for Astrophysics and Cosmology (CAC), University of Nova Gorica, Nova Gorica, Slovenia
\item[$^{77}$] Experimental Particle Physics Department, J.\ Stefan Institute, Ljubljana, Slovenia
\item[$^{78}$] Universidad de Granada and C.A.F.P.E., Granada, Spain
\item[$^{79}$] Instituto Galego de F\'\i{}sica de Altas Enerx\'\i{}as (IGFAE), Universidade de Santiago de Compostela, Santiago de Compostela, Spain
\item[$^{80}$] IMAPP, Radboud University Nijmegen, Nijmegen, The Netherlands
\item[$^{81}$] Nationaal Instituut voor Kernfysica en Hoge Energie Fysica (NIKHEF), Science Park, Amsterdam, The Netherlands
\item[$^{82}$] Stichting Astronomisch Onderzoek in Nederland (ASTRON), Dwingeloo, The Netherlands
\item[$^{83}$] Universiteit van Amsterdam, Faculty of Science, Amsterdam, The Netherlands
\item[$^{84}$] Case Western Reserve University, Cleveland, OH, USA
\item[$^{85}$] Colorado School of Mines, Golden, CO, USA
\item[$^{86}$] Department of Physics and Astronomy, Lehman College, City University of New York, Bronx, NY, USA
\item[$^{87}$] Michigan Technological University, Houghton, MI, USA
\item[$^{88}$] New York University, New York, NY, USA
\item[$^{89}$] University of Chicago, Enrico Fermi Institute, Chicago, IL, USA
\item[$^{90}$] University of Delaware, Department of Physics and Astronomy, Bartol Research Institute, Newark, DE, USA
\item[$^{91}$] University of Wisconsin-Madison, Department of Physics and WIPAC, Madison, WI, USA
\item[] -----
\item[$^{a}$] Louisiana State University, Baton Rouge, LA, USA
\item[$^{b}$] Institut universitaire de France (IUF), France
\item[$^{c}$] also at University of Bucharest, Physics Department, Bucharest, Romania
\item[$^{d}$] now at ECAP, Erlangen, Germany
\item[$^{e}$] School of Physics and Astronomy, University of Leeds, Leeds, United Kingdom
\item[$^{f}$] now at Agenzia Spaziale Italiana (ASI).\ Via del Politecnico 00133, Roma, Italy
\item[$^{g}$] Fermi National Accelerator Laboratory, Fermilab, Batavia, IL, USA
\item[$^{h}$] now at Graduate School of Science, Osaka Metropolitan University, Osaka, Japan
\item[$^{i}$] Max-Planck-Institut f\"ur Radioastronomie, Bonn, Germany
\item[$^{j}$] also at Kapteyn Institute, University of Groningen, Groningen, The Netherlands
\item[$^{k}$] Colorado State University, Fort Collins, CO, USA
\item[$^{l}$] Pennsylvania State University, University Park, PA, USA
\item[$^{m}$] now at Institute of Physics, University of Mainz, Staudinger Weg 7, D-55099 Mainz, Germany
\item[$^{n}$] now at Università di Roma Sapienza and INFN Roma, Italy
\item[$^{o}$] Dept. of Physics and Astronomy, Univ. of Hawai’i, Manoa, USA
\item[$^{p}$] Caltech Jet Propulsion Laboratory, Pasadena, California, USA
\end{description}
}

\begin{abstract}
A dedicated search for upward-going air showers at zenith angles exceeding $110^\circ$ and energies $E>0.1$\,EeV has been performed using the Fluorescence Detector of the Pierre Auger Observatory. The search is motivated by two ``anomalous" radio pulses observed by the ANITA flights I and III which appear inconsistent with the Standard Model of particle physics. Using simulations of both regular cosmic ray showers and upward-going events, a selection procedure has been defined to separate potential upward-going candidate events and the corresponding exposure has been calculated in the energy range [0.1-33] EeV. One event has been found in the search period between 1 Jan 2004 and 31 Dec 2018, consistent with an expected background of $0.27 \pm 0.12$ events from mis-reconstructed cosmic ray showers. 
This translates to an upper bound on the integral flux of $(7.2 \pm 0.2) \times 10^{-21}$~cm$^{-2}$~sr$^{-1}$~y$^{-1}$  and  $(3.6 \pm 0.2) \times 10^{-20}$~cm$^{-2}$~sr$^{-1}$~y$^{-1}$ for an $E^{-1}$ and $E^{-2}$ spectrum, respectively. An upward-going flux of showers normalized to the ANITA observations is shown to predict over 34 events for an $E^{-3}$ spectrum and over 8.1 events for a conservative $E^{-5}$ spectrum,  in strong disagreement with the interpretation of the anomalous events as upward-going showers.
\end{abstract}
\maketitle

The Antarctic Impulsive Transient Antenna (ANITA) instruments, flown on long duration NASA balloons at 30--39~km altitudes above Antarctica, have detected radio pulses that are consistent with coherent emission from Ultra-High-Energy Cosmic Ray (UHECR) air showers. The large horizontal polarization of the pulses is consistent with the geomagnetic effect due to the Earth's magnetic field~\cite{ANITA:2010ect}. 
The few pulses arriving from directions above the horizon are interpreted as direct emission from 
air showers with trajectories that do not intercept the surface of the ice (or Earth, in general), here referred to as ``Earth-missing showers". The majority of the pulses are reflected at the air–ice interface and appear to arrive from the ice surface (below the horizon). They thus display a characteristic polarity inversion. 
In addition, several ``anomalous'' pulses have been reported coming from directions below the horizon~\cite{ANITA:2016vrp,ANITA:2018sgj,ANITA:2020gmv}. These events show strong horizontal polarization, but without the polarity inversion expected for reflected pulses from UHECR showers. The first two such events were detected with the ANITA I and III instruments, respectively at elevations of~$27.4^\circ$~\cite{ANITA:2016vrp} and $35.0^\circ$~\cite{ANITA:2018sgj}, 
corresponding to zenith angles of $\theta = 116.7^\circ$ and $124.5^\circ$ at the intercept of the trajectory with the ice cap.
They could be induced by air showers developing in the upward direction, as could be expected from tau lepton decays produced in UHE tau-neutrino interactions below the surface. However, the direction of the observed pulses implies that the neutrinos would need to travel about 6000-7000 km through the Earth before interacting below the ice surface~\cite{ANITA:2016vrp}. This corresponds to about 8-10 interaction lengths at the required neutrino energy $E_\nu \gtrsim 0.2$~EeV \cite{Connolly:2011vc}, causing severe attenuation and requiring a $\nu_\tau$-flux that should have been observed with IceCube and the Pierre Auger Observatory~\cite{Romero-Wolf:2018zxt,Safa:2019ege,Chipman:2019vjm}, the latter being particularly sensitive to Earth-skimming tau neutrinos~\cite{PierreAuger:2019ens,PierreAuger:2013wqu}. 
An astrophysical explanation of the events under Standard Model (SM) assumptions has also been severely constrained by IceCube \cite{IceCube:2020gbx}.

The shower energy inferred from the pulse amplitudes depends on the altitude, $h$, at which the shower is assumed to start developing 
with respect to the ice level. Simulated showers injected at~$0 < h < 9$~km indicate that the minimum shower energy consistent with the ANITA I (ANITA III) event depends on the unknown shower starting point and is e.g.\ $\sim 0.2$~EeV ($\sim 0.15$~EeV) for showers starting at $h \sim 5$~km ($h>5$~km)~\cite{Romero-Wolf:2018zxt}. 
Explanations based on sub-surface reflections~\cite{Shoemaker:2019xlt} or
coherent transition radiation (TR), expected as an UHECR shower intercepts the ice-air interface, have also been suggested as a possible emission mechanism. TR generated from upward-going showers starting in the ice and intercepting the interface has been ruled out~\cite{Motloch:2016yic}, 
and, similarly, TR due to downward-going UHECR showers intercepting the ice \cite{deVries:2019gzs} is found to have inconsistent polarity according to recent simulations~\cite{Ammerman-Yebra:2023hjp,Ammerman-Yebra:2023xxx}.

Given the difficulties in interpreting the anomalous ANITA events, they have attracted a lot of attention. 
Theoretical interpretations involving physics beyond the SM (BSM) have been put forward invoking new particles that induce upward-going showers in the atmosphere (see e.g.\ \cite{Fox:2018syq,Collins:2018jpg,Heurtier:2019git,Cline:2019snp,Borah:2019ciw,Liang:2021rnv,Reno:2021cdh}). Given the relevance of these observations and their discovery potential, a confirmation or a constraint on upward-going air showers from an independent observation is of particular interest. In this article we search for these showers with the Fluorescence Detector of the Pierre Auger Observatory. Similar approaches using optical telescopes were reported in the context of searches for Earth skimming events induced by interactions of electron neutrinos in the Earth crust \cite{Abbasi:2008hr} or by tau neutrinos producing taus that decay in flight \cite{MAGIC:2018gza} and are also planned for dedicated future experiments, e.g.\ \cite{Otte:2018uxj}.

The Pierre Auger Observatory is the largest cosmic ray detector ever constructed (3000 km$^2$) for UHECR detection above 0.1~EeV~\cite{PierreAuger:2015eyc,PierreAuger:2009esk}.
It was completed in 2008 and combines a Surface Detector (SD), an array of water-Cherenkov detectors to detect the shower front at ground level, and multiple telescopes, known as the Fluorescence Detector (FD), to collect the fluorescence light emitted by nitrogen as the shower front crosses the atmosphere above the SD array. 
Three High Elevation Auger Telescopes (HEAT) were added in 2009 to better record low energy showers~\cite{Mathes:2011zz}. UHECR showers typically develop in the downward direction. 
The timing information from the SD 
combined with the FD data constitutes a {\it hybrid} data set that allows for an improved geometrical reconstruction. The hybrid exposure grows strongly with shower energy, exceeding 300~km$^2$\,sr\,y for $1$~EeV~\cite{PierreAuger:2021ibw,PierreAuger:2010swb}. That of the FD alone has not been investigated before, but can only be expected to be larger.

Upward-going air showers with zenith angles larger than $110^\circ$, as considered here, are unlikely to trigger the SD, so the search presented here uses events having only FD information. 

The standard FD reconstruction is carried out in two stages: First, a geometrical reconstruction of the arrival direction and the impact point of the shower on the ground is made using the timing information from the triggered pixels. Second, the signal traces in the pixels are exploited to obtain the development of the energy deposition as a function of the depth of atmosphere traversed, using the Gaisser-Hillas  profile (GH)~\cite{gaisser:1977ICRC, PierreAuger:2015eyc, PierreAuger:2018gfc}. 

Alternatively, a Global Fit (GF) reconstruction is used to simultaneously find the arrival direction, the impact point, and the GH energy deposition that best fit the complete pixel data~\cite{Novotny:2021lfu}. The analysis takes into account the contribution of scattered Cherenkov light to the signal and it can be used to combine data from several FD locations. As it uses more information, it is more effective in eliminating badly reconstructed events and noise. 
Both the standard and GF reconstructions are applied in either the upward or downward mode within the \Offline{} analysis Framework of the Observatory~\cite{Argiro:2007qg}. 

When the impact point of a shower is in front of a telescope inside the area covered by the SD, as is required for hybrid events, the time sequence of the active pixels clearly defines if it is an upward or a downward-going event. However, the sequence is reversed if the impact point is behind the telescope (see \cite{suppl-material})\nocite{Dawson:2020bkp}. When no SD information is available, the reconstruction may be ambiguous with the fits converging in both the upward and downward solutions, one of them being a mis-reconstruction. 
Ordinary cosmic ray showers, which are reconstructed in the upward direction, constitute an unwanted background.  
This is the case for mis-reconstructed directions and for some Earth-missing showers which are reconstructed with a zenith angle exceeding $90^\circ$, relative to zenith at the center of the SD array. 

An important source of background in data is due to laser pulses. Different types of laser shots are routinely fired across the array from different positions at an average rate of about 150\,Hz to continuously monitor the atmospheric quality and to test and evaluate the performance of both the FD instrument and the reconstruction procedure~\cite{PierreAuger:2015eyc}. 
These laser shots naturally mimic showers traveling in the upward direction and are usually precisely time-tagged so that
they can be easily vetoed.
However, a fraction of order $0.01\%$ of the laser events were not properly labeled and cannot be vetoed. A dedicated effort has been made to identify and discard them by making use of a sample of 10\% of the available FD data up to 31 Dec 2018 ({\it burn sample}). A set of selection cuts, based on the frequency and location of such events, has been defined to clean the burn sample and, presumably, the full data set of all laser events. 

Dedicated simulations of UHECR (background), as well as upward-going showers (signal), have been produced with CONEX~\cite{Bergmann:2006yz}~\footnote{We use Sibyll 2.3c~\cite{Riehn:2017mfm,Ahn:2009wx} and UrQMD 1.3~\cite{Bass:1998ca} respectively for high- and low-energy hadronic interactions.} to optimize the final search in presence of possible background.
For the former, $166\times 10^6$ showers of proton, helium, nitrogen, and iron primaries have been produced in the energy range $0.1$- $100$~EeV.
In the first batch, the UHECR-induced showers were isotropically injected over the surface of a sphere of radius 90\,km, centered on the array. The zenith angles, $\theta$, relative to zenith at the array center, extend to $100^\circ$ to include Earth-missing showers. 
To increase statistics, particularly for inclined events, a second batch of $93\times 10^6$ showers was simulated in the range $60^\circ$-$100^\circ$ (where all the background was found in the first batch, {\it c.f.} \cite{suppl-material}. The energy distribution of the full background sample is weighted to mimic the measured UHECR spectrum~\cite{PierreAuger:2020qqz,PierreAuger:2021ibw}. 

To study the signal, an isotropic distribution of $6 \times 10^7$ upward-going proton showers has been similarly simulated in the range $0.03$-$10$~EeV. 
We note that -- due to shower universality \cite{Ave:2017uiv} -- protons stand in for arbitrary primaries with minimal or no loss of generality, since the first interaction points are directly set and all calculations use the shower energy rather than
the energy of the primary particle.
The combined efficiency for triggering and reconstructing such showers was found to be negligible for energies below $0.03$~EeV. 
The simulated showers are forced to develop at a uniformly distributed altitude above the ground~\footnote{The altitude of the Observatory is taken to be $1400$\,m above sea level.}, $h$, in the range $0<h<9$~km with zenith angles, measured at the exit point on the Earth, between $110^\circ$ and $180^\circ$.
Altitudes $h>9$\,km are not considered because the exposures of Auger and ANITA fall down rapidly here. This is mostly for geometrical reasons (maximum elevation angles of the Auger FD telescopes and a narrowing Cherenkov cone in case of ANITA).
The flat distribution of shower starting points used in the simulations is no restriction, as any distribution can be generated from this by applying corresponding weights to the altitude bins (see below).
The ground impact (exit) points have been sampled in a square area of $100 \times 100$~km$^2$ centered on the SD array. 
This area extends up to $\sim 20$~km behind each FD site to include simulated trajectories with exit points behind the field-of-view of a telescope \cite{suppl-material}. 
Additionally, to increase statistics in the low energy region, $5 \times 10^6$ proton showers have been simulated between $0.1$ and $0.3$~EeV with impact points contained in a circle with a radius increasing with energy from 12~km to 23~km around HEAT but otherwise with the same distributions. 

The simulation of the FD signals and trigger, and the subsequent event reconstruction, are done within \Offline{}
to study the performance of the reconstruction algorithms. 
The reconstructed zenith angles correlate well with their true value. However, as no cuts targeted to directional reconstruction precision were applied, a tail in $\theta_{\rm rec}-\theta_{\rm sim}$ is present leading to a $68\%$ central interval of $[-1.1^\circ,11^\circ]$ (See \cite{suppl-material} for further details).

\begin{figure}[tb]
\centering
\includegraphics[scale=0.45]{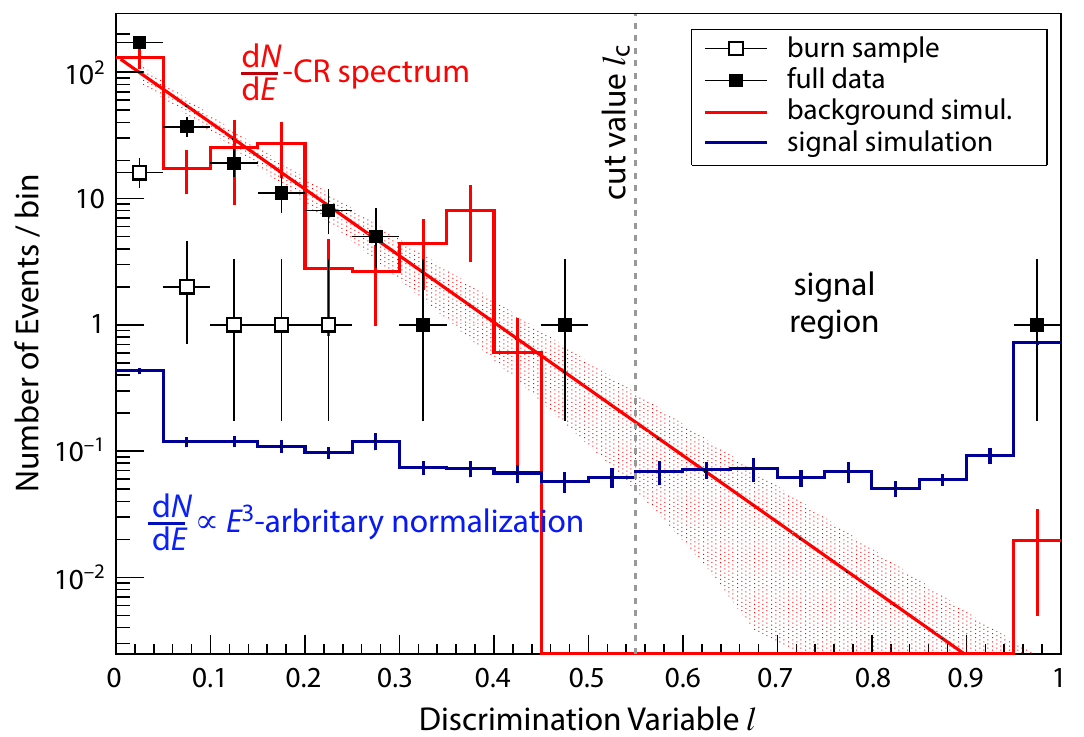}
\caption{Distributions of the discriminating variable $l$, as defined in Eq.\,(\ref{eq:discrimination-variable}), for \textit{i}) a simulated isotropic background weighted and normalized with the measured UHECR spectrum in the energy range $10^{17}$~eV to $10^{20}$~eV~\cite{PierreAuger:2020qqz} (red histogram with an exponential fit and its uncertainty band); \textit{ii})  the signal simulation with energy $10^{16.6}$~eV to $10^{18.5}$~eV weighted with an $E^{-3}$ spectrum and arbitrarily normalized to one event (blue histogram); and \textit{iii})  the data distributions, both for the 10\% burn sample and the full data set (open and filled symbols). The cut value $l_c$ discriminating signal and background is indicated by the vertical dashed line.}
\label{fig:lDist}
\end{figure}

The selection of candidates compatible with upward-going showers that exit the Earth's surface was performed making use of simulations to reduce the large background to a minimal level. After deciding the entire selection strategy, it was applied to the aforementioned 10\% burn sample for verification (see Fig.\,\ref{fig:lDist}) before it was finally applied blindly to the full FD data set ($7.6 \times 10^6$  events). 
In the first step, the aforementioned laser cuts reduce the data sample to $4.7\times 10^6$ events.
To guarantee a minimum data quality~\cite{PierreAuger:2010swb}, only time periods with a clean atmosphere and low cloud coverage \cite{PierreAuger:2014sui}, and only events with at least six camera pixels contributing to the time-geometry fit of the shower axis \cite{PierreAuger:2009esk} are considered, leaving $\sim 600$k events. 
Out of these, 165k events can be reconstructed as upward-going in a simple time-geometry fit \cite{PierreAuger:2009esk}, if only the time-sequence and pointing direction of the triggered pixels are considered. 

The GF reconstruction is used to check whether the pixel intensity is consistent with a shower-like $dE/dX$-profile eliminating many events where the dominant signal is from Cherenkov light. Because of their time-compressed structure they are 
mis-reconstructed as upward-going by the simple time fit \textit{c.f.} \cite{suppl-material}. Only 2774 events survive this step. The GF is then also applied in the downward mode and it is found that the majority of the selected events allow both upward and downward reconstructions.  
Only 986 events are left when two further quality constraints are applied ensuring that the interval of atmospheric slant depth, over which the shower profile is observed, exceeds 80~g\,cm$^{-2}$ and the reconstructed shower maximum is above ground. 
To eliminate background from mis-reconstructed and Earth-missing showers, a cut of $\theta > 110^\circ$ is applied leaving 928 events.
Finally, events
with $\chi^2_{\rm up} \geq 1.2~\chi^2_{\rm down}$ are removed as the downward reconstruction is clearly preferred over the upward reconstruction \footnote{The final search criterion, discussed below, is based on comparing the more precise likelihood ratios.}, reducing the sample to 255 events (\textit{c.f.} Fig.~\ref{fig:lDist}). The effects of cuts on data and simulation are compared in the SM. 

A search criterion is finally needed to optimize discrimination between upward-going showers (signal) and cosmic-ray events (background). For convenience, we use a function of the logarithm of the likelihood ratio of the upward and downward modes, $L_{\mathrm{up}}/L_{\mathrm{down}}$,

\begin{equation}
l = \frac{\arctan{\{\ln{[\max(L_{\mathrm{up}}, L_{\mathrm{down}})/L_{\mathrm{down}}]}\cdot \zeta\}}}{\pi/2}
\label{eq:discrimination-variable}
\end{equation}

\noindent such that the discrimination variable $l$ ranges from $0\leq l \leq 1$, with larger values reflecting larger ratios $L_{\mathrm{up}}/L_{\mathrm{down}}$. When $L_{\mathrm{down}}$ exceeds $L_{\mathrm{up}}$, its value is $l=0$. We note that, occasionally, some simulated cosmic ray showers can only be reconstructed as upward-going. In such a case, the event is assigned a value of $l=1$. Finally, the scale parameter $\zeta$ in Eq.\,\ref{eq:discrimination-variable} is 
chosen such that the signal events uniformly cover the range [0,1].
Figure\,\ref{fig:lDist} presents the resulting $l$-distributions for both the UHECR background (red) and the upward-going signal simulations (blue). 
While the signal distribution is rather flat, the background drops down by about 4 orders of magnitude across the range of $l$.

A final cut $l>l_c$ is applied to minimize the UHECR background while keeping a sufficiently large fraction of the signal. 
To optimize $l_c$, the $l$-distribution of the simulated background has been fitted using several different trial functions. For each fitted distribution, an optimal cut value, $l_c$, is chosen by performing a scan on $l$ to find the value that minimizes the upper limit obtained for the integral flux of upward-going showers, as discussed below. 
The flux limit to be minimized in the $l_c$ scan is obtained using an exposure weighted over energy with power laws  $E^{-1}$ and $E^{-2}$, assuming uniform distribution in $h$
and fixing the number of observed events $n_{\mathrm{obs}}$ to be equal to the number of events expected from the cosmic ray background $n_{\mathrm{bkg}}$ for a given value of $l_c$.  
With these assumptions, the optimal value is found at $l_c = 0.55$ for both considered spectral indices~\cite{PierreAuger:2021gci}. Above this value, the expected background for the full data set 
is $n_{\mathrm{bkg}} = 0.27 \pm 0.12$. 
(We note that all simulated background events passing the $l$-cut in Fig.\,\ref{fig:lDist} are found at $l=1$. They have simulated zenith angles close to $90^\circ$ and are mis-reconstructed with $\theta_{\rm rec}>110^\circ$. See \cite{suppl-material} for further details.)
Different parameterizations of the fit to the UHECR background affect the upper limits within 10\,\%. This is included in the quoted uncertainty of the expected background, $n_{\mathrm{bkg}}$. 

Once a value of $l_c$ is chosen, the full selection and search procedure is completely defined. The sequence of selection cuts is performed on the simulated showers in an identical way as on the data, and it is shown to have similar effects for both \cite{suppl-material}.  

\begin{figure}[tb]
\includegraphics[scale=0.43]{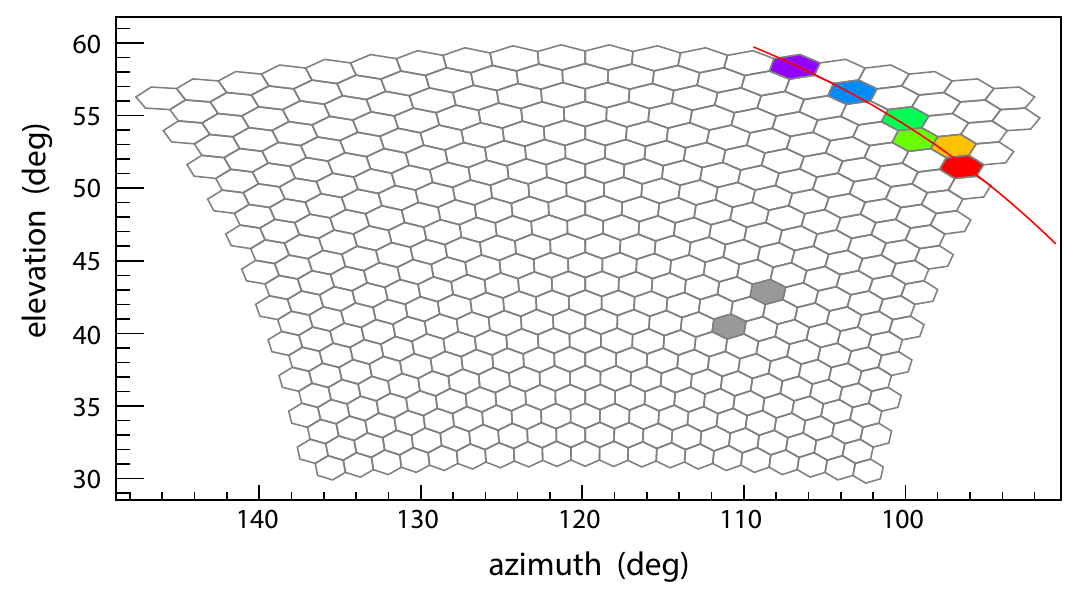}
\includegraphics[scale=0.46]{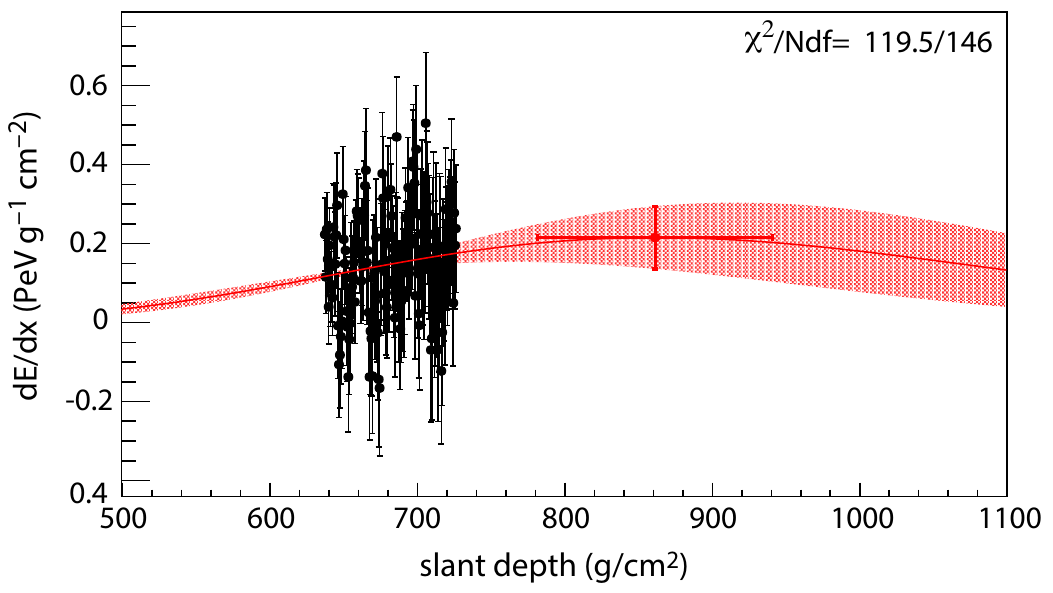} 
\caption{Remaining event after the selection and search procedure. The top panel shows the triggered pixels of the camera, the earliest one in purple and the last one in red.
The bottom plot shows the reconstructed profile that has been fit with the GF reconstruction in the upward-going mode.
The time evolution of the signal across a pixel is divided into 50 ns bins to give information from different atmospheric depths.
}
\label{fig:bkg.event}
\end{figure}

The distribution of events from the full data set passing the selection criteria agrees well with both the burn sample and the background simulations (\textit{c.f.}\ Fig.~\ref{fig:lDist} and \cite{suppl-material}). After unblinding, one candidate event with {$l=1$} was found, consistent with background expectations, with key features depicted in Fig.\,\ref{fig:bkg.event}. Its FD image sweeps a small portion in the top corner of a HEAT camera, triggering only six pixels, i.e.\ the minimum defined in the quality cuts. The reconstruction quality of such events is moderate and in consequence, several poorly imaged events can be found in the background simulations ({\it c.f.} \cite{suppl-material}).

Applying the event selection criteria  to the simulated signal allows us to calculate the effective area of the detector for a flux of upward-going air showers as a function of shower energy, $E$, and starting altitude, $h$. 

An integral flux limit is then obtained by taking the ratio of the maximum number of events allowed by the search to the exposure for selected upcoming showers, ${\langle \cal E \rangle} (E>E_0)$, which is weighted with a given energy spectrum. The maximum number of allowed events is taken to be the Rolke limit~\cite{Rolke:2004mj} at 95\% CL, $N^{95\%}(n_{\mathrm{bkg}},n_{\mathrm{obs}})$, where $n_{\mathrm{obs}}$ is the actual number of events observed after unblinding and $n_{\mathrm{bkg}}$ the expected number of events for the cosmic ray background above the specified $l_c$ cut:  

\begin{equation}
    F^{95\%} (E > E_{0}) = \frac
    {N^{95\%}(n_{\mathrm{bkg}},n_{\mathrm{obs}})}
    {{\langle \cal E \rangle} (E>E_0)}\, .
    \label{upperlimit}
\end{equation}

Injecting $n_{\rm obs}$ and $n_{\rm bkg}$ from above into this equation  and assuming power law spectra $E^{-1}$ and $E^{-2}$ in $E \in [0.1,33]$\,EeV and a distribution uniform in $h$ and isotropic in $\theta$, we find integral flux upper limits at $(7.2 \pm 0.2) \times 10^{-21}$~cm$^{-2}$~sr$^{-1}$~y$^{-1}$ and $(3.6 \pm 0.2) \times 10^{-20}$~cm$^{-2}$~sr$^{-1}$~y$^{-1}$, respectively.
Preliminary results based on lower Monte Carlo statistics and using different energy ranges have been presented in \cite{PierreAuger:2021gci,PierreAuger:2023elf}.

\begin{figure}[tbh!]
\centering
\includegraphics[width=0.48\textwidth]{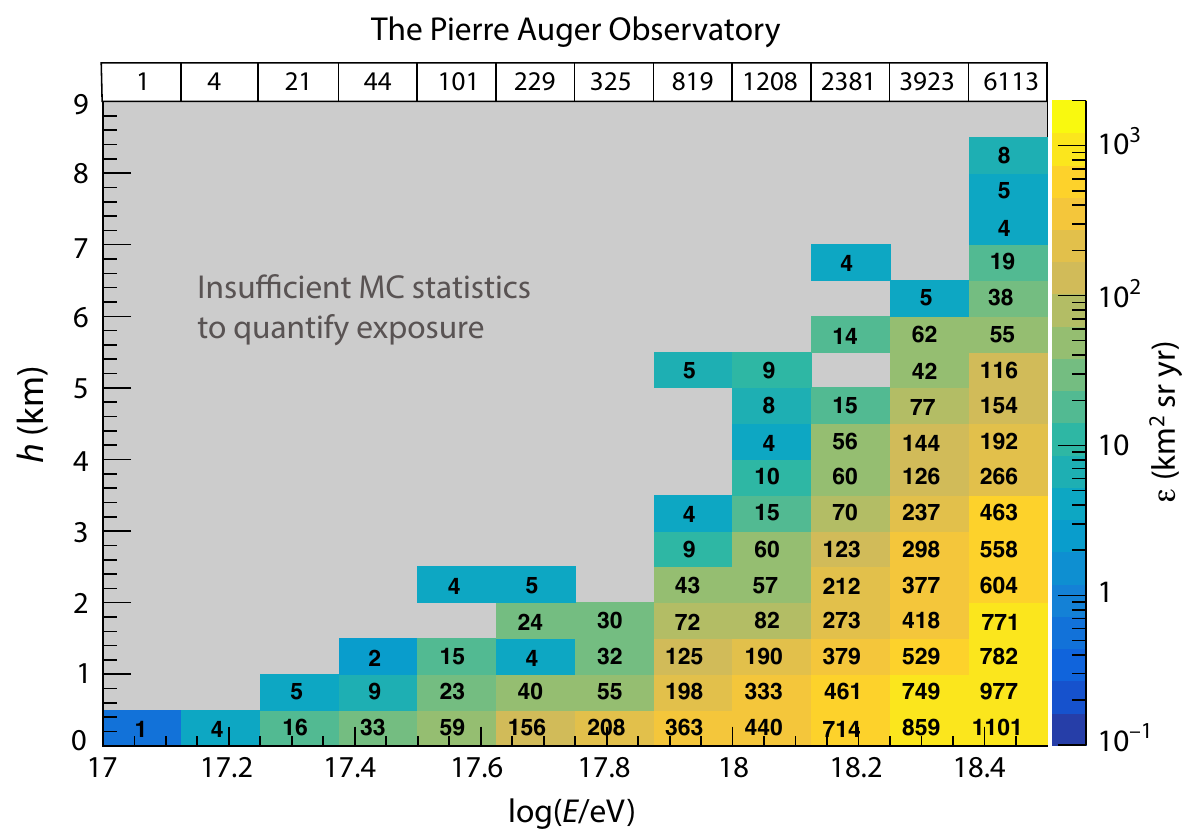}
\includegraphics[width=0.48\textwidth]{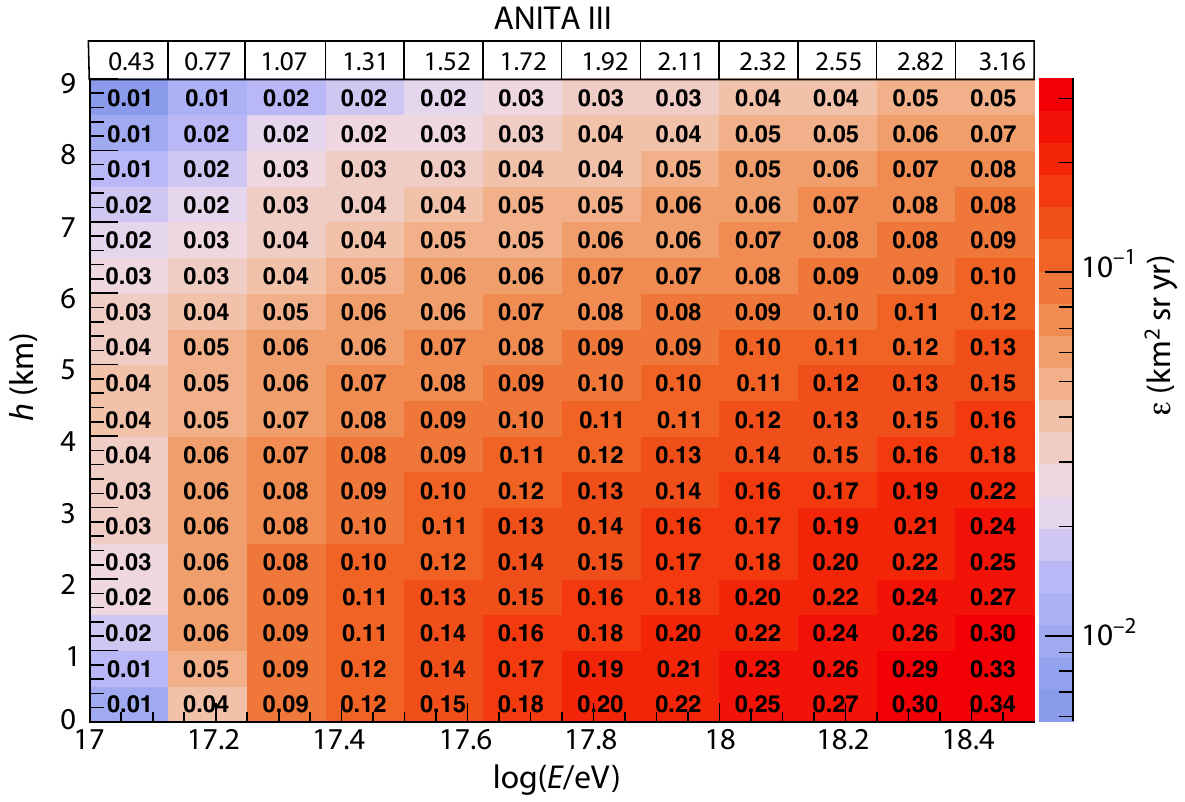}
\caption{Exposure of the Auger Observatory (top) and the ANITA III flight (bottom) as a function of shower energy, $E$, and injection altitude, $h$, integrated over the zenith angle range $110^\circ \leq \theta \le 130^\circ$, for an 
isotropic distribution of arrival directions. The gray area indicates insufficient statistics. In the white cells on top of the exposure plots, we display the sums of the $h$-bins to facilitate comparison (see text).}
\label{fig:ddexposure}
\end{figure}

To relate the non-observation of upward-going showers at the Auger Observatory to the observation of the anomalous events in ANITA, we calculate 2D exposure maps for both observatories as a function of shower energy and shower starting point in the atmosphere. This is done for three zenith angle ranges: $110^\circ-130^\circ$, $130^\circ-145^\circ$ and $145^\circ-180^\circ$ and is depicted in Fig.~\ref{fig:ddexposure}  (top) for the first angular bin which largely overlaps with the ANITA anomalous events.
The acceptance of the ANITA detector to upward-going showers has been calculated using an analytical approach
integrating the surface area and solid angle over a spherical surface concentric with the Earth at altitude $h$. The accepted solid angle is approximated by the maximum and minimum off-axis angles (of the radio pulse relative to the shower axis) within which the recorded peak amplitude is seen at the detector above a fixed value. These angles are obtained with a set of proton simulations at $0.1$~EeV starting at $h$ between 0 and 9 km (above the ice surface) for zenith angles in the range $90^\circ$-$130^\circ$~\cite{Romero-Wolf:2018zxt}. The pulses are assumed to scale linearly with shower energy. Threshold values of 446~\textmu{V}\,m$^{-1}$ and 284~\textmu{V}\,m$^{-1}$ have been used respectively for ANITA I and III instruments~\cite{Romero-Wolf:2018zxt}. 
The exposure for the $110^\circ-130^\circ$ range in zenith angle has been obtained by multiplying the effective area by the effective flight time, which is 17 (7) days for the ANITA I (III) flight~\cite{schoorlemmer2016energy,ANITA:2018sgj}. 
It is displayed in Fig.~\ref{fig:ddexposure} (bottom) for ANITA III and has been cross-checked modifying a Monte Carlo simulation developed for the calculation of the exposure to tau neutrinos in the ANITA IV flight~\cite{ANITA:2021xxh}~\footnote{Several authors of this article are members of the ANITA collaboration and have been involved in these calculations}. 
The two exposures increase with the energy of upward-going showers. That of the Auger Observatory can only be calculated for showers starting at low altitudes, $h$, because it falls very rapidly with $h$, and the required simulation statistics become unfeasible to produce, particularly at lower energies. 
The sums of the exposure bins in $h$ illustrate that the sensitivity of the Auger Observatory exceeds that of the ANITA-III flight by factors rising with energy from about 2 to 2000 for a uniform $h$-distribution.   

Assuming that the anomalous ANITA events are indeed produced by upward-going showers, we can then calculate the number of upward-going showers expected in the data of the Auger Observatory by convolving a given spectral flux and an $h$-distribution of the showers, with the two 2D exposure maps.
Three power-law spectra, $E^{-\gamma}$, with $\gamma = 2,3,5$ have been assumed, and for each case, we consider both a uniform distribution in $h$ and that expected for exiting taus that decay in the atmosphere. In the latter case, expected from tau neutrino interactions and in many of the proposed BSM scenarios, the energy $E$ is that of the tau leptons. 
The starting altitude of the shower strongly depends on
zenith angle and tau energy. The required distribution in $E$ and $h$ is obtained from a convolution of the tau flux, the decay-length distribution and the 
distribution of shower energy for all tau decays as obtained with TAUOLA~\cite{Chrzaszcz:2016fte}, with an average of $\sim 50\,\%$ of the tau energy. 

A given energy spectrum can be normalized to the anomalous observations demanding one expected event after folding with the ANITA I or III exposures.
Normalizing to ANITA I (III) observation under the assumption of an $E^{-3}$ spectrum and a uniform $h$-distribution, we expect 59 (69) events at the Auger Observatory.  Using an $h$-distribution compatible with tau decay reduces expectations to 37 (34) events. 
Assuming a very conservative spectrum, $E^{-5}$, to the event of flight I (III) results in 11.7 (8.1) expected events for a uniform $h$-distribution and 18 (11) events for the $h$-distribution expected from tau decay.  These numbers are to be compared with one observed event compatible with background.
We note that, given a spectral index, the expected number of events obtained using normalizations from the two flights are similar to one another. The significantly lower signal threshold of the ANITA III instrument is compensated by a lower effective flight duration. 

The results of this search do not support the interpretation that the anomalous pulses detected during the ANITA I and III flights were caused by  air showers sourced from particle interactions or decays, such as by decays of upward going taus, the latter being the basis of proposed explanations based on physics beyond the SM. 
When comparing the results of this study with those reported by ANITA, we note that the exposures of the two detectors have very different dependence on shower altitude. The Auger exposure in the gray area of Fig.\,\ref{fig:ddexposure} (top) is generally below a few km$^2$\,sr\,yr and was set to zero because of insufficient Monte Carlo statistics. 
However, the large values in the bins at lower altitude overcompensate this effect, unless 
the showers were distributed exclusively in this yet uncovered region, and/or with mechanisms producing showers with very different longitudinal profiles.
No simple mechanism can be anticipated to produce such distributions. It can thus be argued that the upward-going shower explanation is ruled out for a diffuse flux, unless the distribution of the starting altitude of the showers was shaped such that they would only start several kilometers above ground or the shower profiles had unusual shapes. Both would be inconsistent with showers originating from known particle decays or interactions.

\section*{Acknowledgments}

\begin{sloppypar}
The successful installation, commissioning, and operation of the Pierre
Auger Observatory would not have been possible without the strong
commitment and effort from the technical and administrative staff in
Malarg\"ue. We are very grateful to the following agencies and
organizations for financial support:
\end{sloppypar}

\begin{sloppypar}
Argentina -- Comisi\'on Nacional de Energ\'\i{}a At\'omica; Agencia Nacional de
Promoci\'on Cient\'\i{}fica y Tecnol\'ogica (ANPCyT); Consejo Nacional de
Investigaciones Cient\'\i{}ficas y T\'ecnicas (CONICET); Gobierno de la
Provincia de Mendoza; Municipalidad de Malarg\"ue; NDM Holdings and Valle
Las Le\~nas; in gratitude for their continuing cooperation over land
access; Australia -- the Australian Research Council; Belgium -- Fonds
de la Recherche Scientifique (FNRS); Research Foundation Flanders (FWO),
Marie Curie Action of the European Union Grant No.~101107047; Brazil --
Conselho Nacional de Desenvolvimento Cient\'\i{}fico e Tecnol\'ogico (CNPq);
Financiadora de Estudos e Projetos (FINEP); Funda\c{c}\~ao de Amparo \`a
Pesquisa do Estado de Rio de Janeiro (FAPERJ); S\~ao Paulo Research
Foundation (FAPESP) Grants No.~2019/10151-2, No.~2010/07359-6 and
No.~1999/05404-3; Minist\'erio da Ci\^encia, Tecnologia, Inova\c{c}\~oes e
Comunica\c{c}\~oes (MCTIC); Czech Republic -- GACR 24-13049S, CAS LQ100102401,
MEYS LM2023032, CZ.02.1.01/0.0/0.0/16{\textunderscore}013/0001402,
CZ.02.1.01/0.0/0.0/18{\textunderscore}046/0016010 and
CZ.02.1.01/0.0/0.0/17{\textunderscore}049/0008422 and CZ.02.01.01/00/22{\textunderscore}008/0004632;
France -- Centre de Calcul IN2P3/CNRS; Centre National de la Recherche
Scientifique (CNRS); Conseil R\'egional Ile-de-France; D\'epartement
Physique Nucl\'eaire et Corpusculaire (PNC-IN2P3/CNRS); D\'epartement
Sciences de l'Univers (SDU-INSU/CNRS); Institut Lagrange de Paris (ILP)
Grant No.~LABEX ANR-10-LABX-63 within the Investissements d'Avenir
Programme Grant No.~ANR-11-IDEX-0004-02; Germany -- Bundesministerium
f\"ur Bildung und Forschung (BMBF); Deutsche Forschungsgemeinschaft (DFG);
Finanzministerium Baden-W\"urttemberg; Helmholtz Alliance for
Astroparticle Physics (HAP); Helmholtz-Gemeinschaft Deutscher
Forschungszentren (HGF); Ministerium f\"ur Kultur und Wissenschaft des
Landes Nordrhein-Westfalen; Ministerium f\"ur Wissenschaft, Forschung und
Kunst des Landes Baden-W\"urttemberg; Italy -- Istituto Nazionale di
Fisica Nucleare (INFN); Istituto Nazionale di Astrofisica (INAF);
Ministero dell'Universit\`a e della Ricerca (MUR); CETEMPS Center of
Excellence; Ministero degli Affari Esteri (MAE), ICSC Centro Nazionale
di Ricerca in High Performance Computing, Big Data and Quantum
Computing, funded by European Union NextGenerationEU, reference code
CN{\textunderscore}00000013; M\'exico -- Consejo Nacional de Ciencia y Tecnolog\'\i{}a
(CONACYT) No.~167733; Universidad Nacional Aut\'onoma de M\'exico (UNAM);
PAPIIT DGAPA-UNAM; The Netherlands -- Ministry of Education, Culture and
Science; Netherlands Organisation for Scientific Research (NWO); Dutch
national e-infrastructure with the support of SURF Cooperative; Poland
-- Ministry of Education and Science, grants No.~DIR/WK/2018/11 and
2022/WK/12; National Science Centre, grants No.~2016/22/M/ST9/00198,
2016/23/B/ST9/01635, 2020/39/B/ST9/01398, and 2022/45/B/ST9/02163;
Portugal -- Portuguese national funds and FEDER funds within Programa
Operacional Factores de Competitividade through Funda\c{c}\~ao para a Ci\^encia
e a Tecnologia (COMPETE); Romania -- Ministry of Research, Innovation
and Digitization, CNCS-UEFISCDI, contract no.~30N/2023 under Romanian
National Core Program LAPLAS VII, grant no.~PN 23 21 01 02 and project
number PN-III-P1-1.1-TE-2021-0924/TE57/2022, within PNCDI III; Slovenia
-- Slovenian Research Agency, grants P1-0031, P1-0385, I0-0033, N1-0111;
Spain -– Ministerio de Ciencia e Innovaci\'on/Agencia Estatal de Investigaci\'on
(PID2019-105544GB-I00, PID2022-140510NB-I00 and RYC2019-027017-I), 
Xunta de Galicia (CIGUS Network of Research Centers,
Consolidaci\'on 2021 GRC GI-2033, ED431C-2021/22 and 2022 ED431F-2022/15),
Junta de Andaluc\'\i a (SOMM17/6104/UGR and P18-FR-4314),and the European Union (Marie Skłodowska-Curie 101065027 and ERDF); USA -- Department of
Energy, Contracts No.~DE-AC02-07CH11359, No.~DE-FR02-04ER41300,
No.~DE-FG02-99ER41107 and No.~DE-SC0011689; National Science Foundation,
Grant No.~0450696; The Grainger Foundation; Marie Curie-IRSES/EPLANET;
European Particle Physics Latin American Network; and UNESCO.
\end{sloppypar}

\bibliographystyle{apsrev4-2}
\bibliography{upgoing-search}

\onecolumngrid\newpage
\renewcommand*{\thefootnote}{\fnsymbol{footnote}}

\begin{center}
\large{\bf Supplemental Material for:}\\[1ex]
\large{\bf ``A search for the anomalous ANITA events using the Pierre Auger Observatory''}\footnote[1]{email: spokespersons@auger.org}
\vspace*{3mm}
\newline
\normalsize (The Pierre Auger Collaboration)
\email{spokespersons@auger.org}
\end{center}

\setcounter{figure}{0}

\twocolumngrid

The search for air showers consistent with the anomalous events detected by the ANITA instruments demands identifying shower trajectories that cross the surface of the Earth and develop in the atmosphere in the upward direction. We refer to such showers as {\em upward-going} and their zenith angle, defined at the the point of the trajectory where it exits Earth's surface, is greater than $90^\circ$, measured from zenith. These showers can start developing at any point along the trajectory and the shower maximum, conventionally measured along the shower direction starting at the exit point of the Earth, can reach large values. If the shower starts high up in the atmosphere and is not very inclined, the shower may not reach its shower maximum before exiting the atmosphere. The search discussed here demands the reconstructed shower maximum to be above ground, though not necessarily in the field-of-view (FoV) of the telescopes, i.e.\ only part of the rising tail 
of the shower profile needs to be identified. Cosmic rays with the same trajectories always develop in the downward direction and their zenith angle, defined at the entry point into the Earth, is always less than $90^\circ$. Their depth of shower maximum is conventionally measured from the top of the atmosphere in the downward direction. 

The challenge of the analysis is twofold: one needs a) to identify and reject unwanted background arising from mis-reconstructed UHECR-induced air showers or from artificial sources, such a laser shots that are fired across the Observatory for atmospheric monitoring and b) to quantify the efficiency of properly identifying genuine upward-going air showers.

\begin{figure}[h]
\centerline{\includegraphics[scale=0.45]{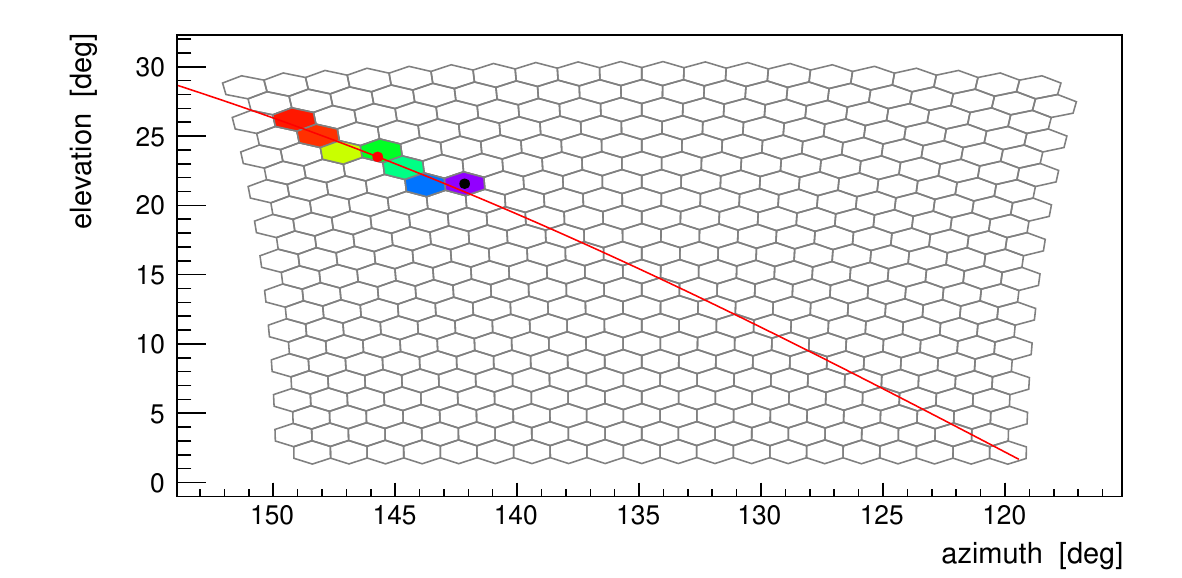}}
\centerline{\includegraphics[scale=0.45]{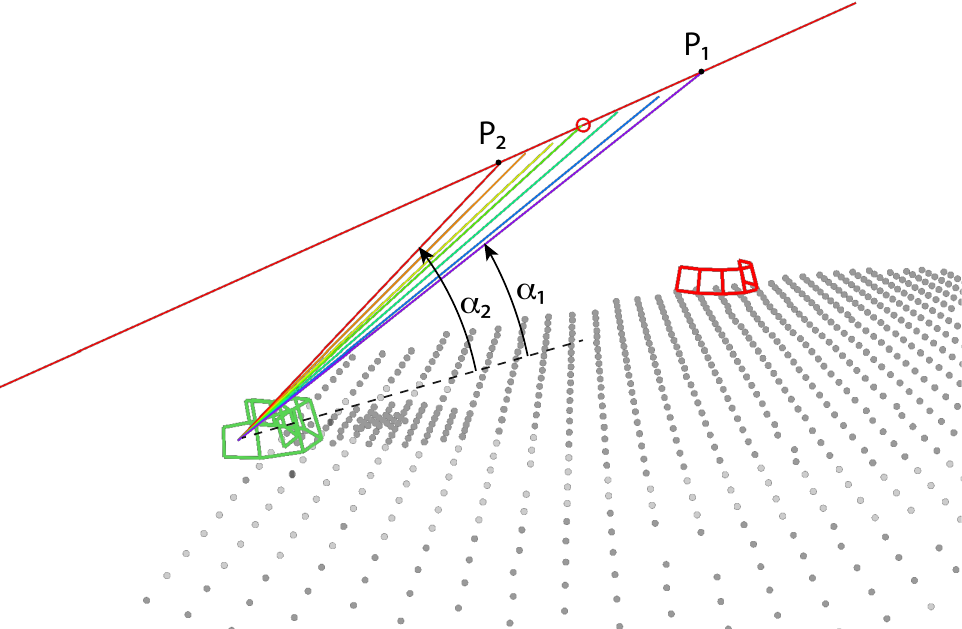}}
\caption{
Example of a measured downward-going cosmic ray event with impact point behind a telescope building. The camera view is shown on top and a 3d view of the shower geometry is shown at the bottom. The colors indicate the time ordering of triggered pixels (purple first, red last). The light from $P_1$ at the low elevation angle $\alpha_1$ arrives at the FD-telescope earlier than that from $P_2$ at the higher elevation angle $\alpha_2$ so that the trace left by the pixels in the camera appears "upward-going" as would correspond to a genuine upward-going shower with its exit point in front of the telescope.
}
\label{fig:backlander}
\end{figure}

An example of potential UHECR-induced background is presented in Fig.\,\ref{fig:backlander}. It shows an air shower which has its impact point located behind the observing telescope, i.e.\ in an area where no surface detector stations are located to detect the shower front. Such trajectories can be viewed with the Fluorescence Detector (FD) if their zenith angle is greater than $30^\circ$ for the HEAT telescope and above $60^\circ$ for the remaining telescopes \cite{PierreAuger:2009esk}. 
The signal coming from the beginning of shower development (purple pixel and light path) has a lower elevation angle and arrives earlier than that from the end of the shower (red pixel and light path) 
and the image in the camera moves upwards as an upward-going shower would appear if its exit point was in front of the telescope.  Such events need to be discriminated and this can be done with a time fit, provided that the visible angular track is sufficiently long. However, when the shower points to the detector close to or within the Cherenkov angle, the time sequence of the triggered pixels is compressed and reconstruction with a simple time fit is problematic. The more restrictive Global Fit (GF), demanding the shower profile to spread in slant depth as can be expected for showers involving Standard Model particles, gets rid of most of the mis-reconstructed events obtained with a time fit alone.

Earth-missing showers are another class of events: they have trajectories that skim the atmosphere with near horizontal directions without intercepting the Earth's surface. Defining the zenith angle of such trajectories requires a geographic location at which a tangent plane to the Earth fixes a coordinate system to measure the zenith angle. 
Technically, the definition used for the zenith angle, $\theta$, of standard events at Auger uses the tangent plane at the intercept of the shower axis with the ground surface (WGS84 ellipsoid going through it). For Earth-missing showers, we have defined the zenith angle relative to a tangent plane at the array center.
For such showers, $\theta$ can be reconstructed below and above $90^\circ$ (typically $87^\circ \lesssim \theta \lesssim 93^\circ$), depending on geometry. 
Naturally, the more inclined the showers are, the easier it becomes for them to be mis-reconstructed in the upward-going direction and also Earth-missing showers are by definition very inclined. 
A cut on $\theta_{\rm rec} > 110^\circ$ has been enforced to reduce this background while still covering well the zenith angles of the two anomalous ANITA events. 

In this Supplemental Material (SM), we discuss the main sources of background due to mis-reconstructed UHECR-induced showers and we demonstrate that the successive quality cuts, optimized with simulations and verified using a burn sample of 10\% of the data, reduce the background very effectively and affect measured data and simulations in a consistent way, thereby giving confidence in the analysis. Having verified this consistency, we demonstrate that the Auger FD telescopes allow the identification and reconstruction of genuine upward-going showers.

\section{Background from specific UHECR event geometries}

When searching for upward-going air showers with the FD, cosmic rays arriving near-horizontal and at highly-inclined downward-going directions represent an inherent source of background. Their contribution has been simulated with UHECRs, injected isotropically over
the surface of a sphere of radius 90 km, centered on the array.

The background then arises due to mis-reconstructed zenith angles, occurring particularly for very inclined and Earth-missing showers. $95 \times 10^6$ events have been simulated with $\theta_{\rm sim}>80^\circ$ (out of a total of $260\times 10^6$) and, after all selection cuts except for the cut on the $S/B$ discrimination variable, $l$, only 124 events were mis-reconstructed with $\theta_{\rm rec} > 110^\circ$.
Applying the cut $l>0.55$ (See Eq.\ 1 in the main article) that finally discriminates signal from background, only five events remain. Their simulated and reconstructed zenith angles are:\\[1ex]
\hspace*{1,5cm}
$\theta_{\rm sim} = 89.6^\circ  \; ; \; \theta_{\rm rec} = 116.7^\circ$ \\
\hspace*{1,5cm}
$\theta_{\rm sim} = 90.7^\circ  \; ; \; \theta_{\rm rec} = 142.1^\circ$ \\
\hspace*{1,5cm}
$\theta_{\rm sim} = 90.8^\circ \; ; \;
\theta_{\rm rec} = 115.6^\circ $ \\
\hspace*{1,5cm}
$\theta_{\rm sim} = 91.9^\circ \; ; \;
\theta_{\rm rec} = 112.9^\circ $ \\
\hspace*{1,5cm}
$\theta_{\rm sim} = 93.0^\circ \; ; \;
\theta_{\rm rec} = 136.2^\circ $.\\[.2ex]

Interestingly, all of these events are near horizontal showers, one slightly downward-going and four slightly upward-going. All of them could only be reconstructed as upward-going by the GF so that they were assigned $l=1$. 
It is these five events that constitute (after proper weighting) the expected background of $0.27 \pm 0.12$ events in the background histogram of Fig.\,1 in the main article.

\begin{figure}[t]
\centerline{\includegraphics[scale=0.45]{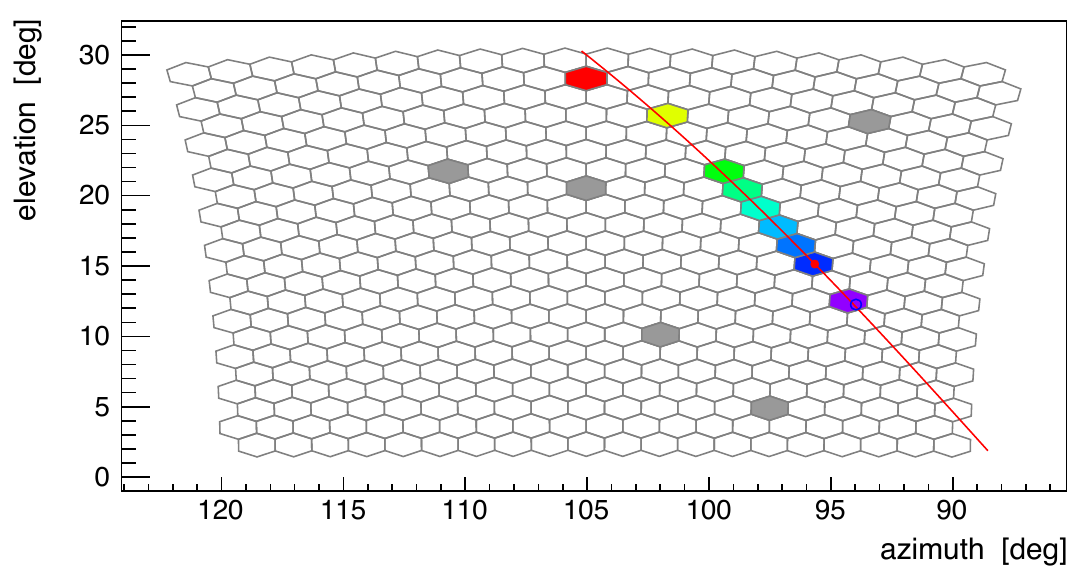}}
\centerline{\includegraphics[scale=0.45]{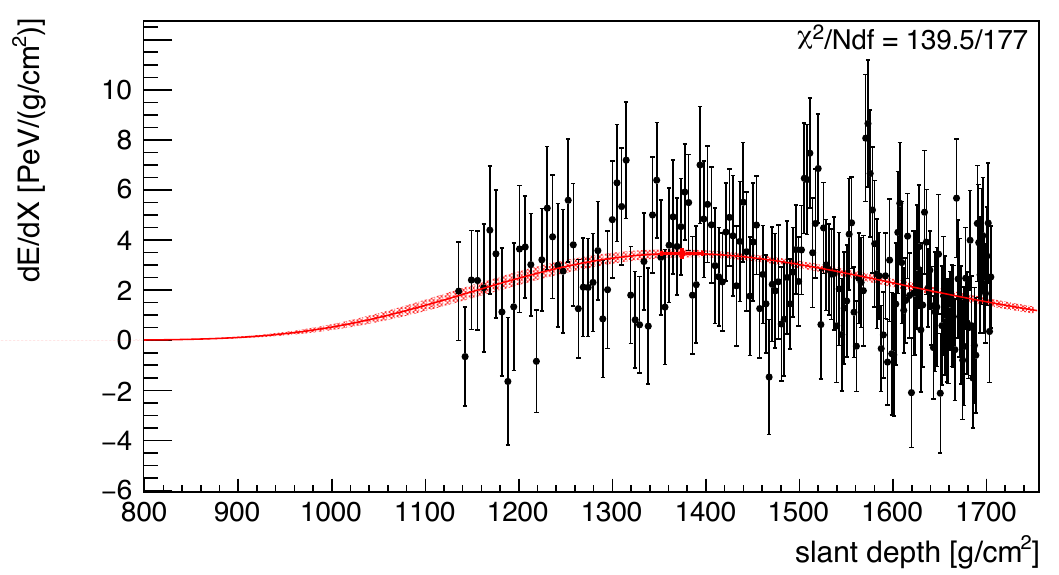}}
\caption{Simulated near horizontal UHECR shower ($\theta_{\rm sim}=89.6^\circ$) that is mis-reconstructed as upward-going and part of the background with a discrimination variable $l=1$ (\textit{c.f.} Eq.\ 1 in main text). The top plot shows a camera view of the event. The colors indicate again the time ordering of triggered pixels (purple first, red last). Isolated pixels shown in gray color are noise pixels that are not used for the reconstruction. The bottom plot shows the mis-reconstructed profile of deposited energy for the upward-going geometry using the standard methods to convert the amount of light per time-bin into energy loss, $dE/dX$, per slant depth, $X$ \cite{PierreAuger:2014sui}.
}
\label{fig:backl=1}
\end{figure}

An example of such a shower, the first of the five events listed above, is shown in Fig.\,\ref{fig:backl=1}. It is simulated with an energy of 6.7 EeV and develops above the Auger Observatory, traveling almost parallel to the ground and diagonally through the FoV of the camera. 
The simulated depth of shower maximum is at 682.5\,g\,cm$^{-2}$. The mis-reconstructed profile, shown in Fig.\,\ref{fig:backl=1}, has a shower maximum at 1375\,g\,cm$^{-2}$ (measured relative to the exit point of the trajectory, as it is interpreted as an upward going shower).

\begin{figure}[tbh]
\centerline{\includegraphics[scale=0.45]{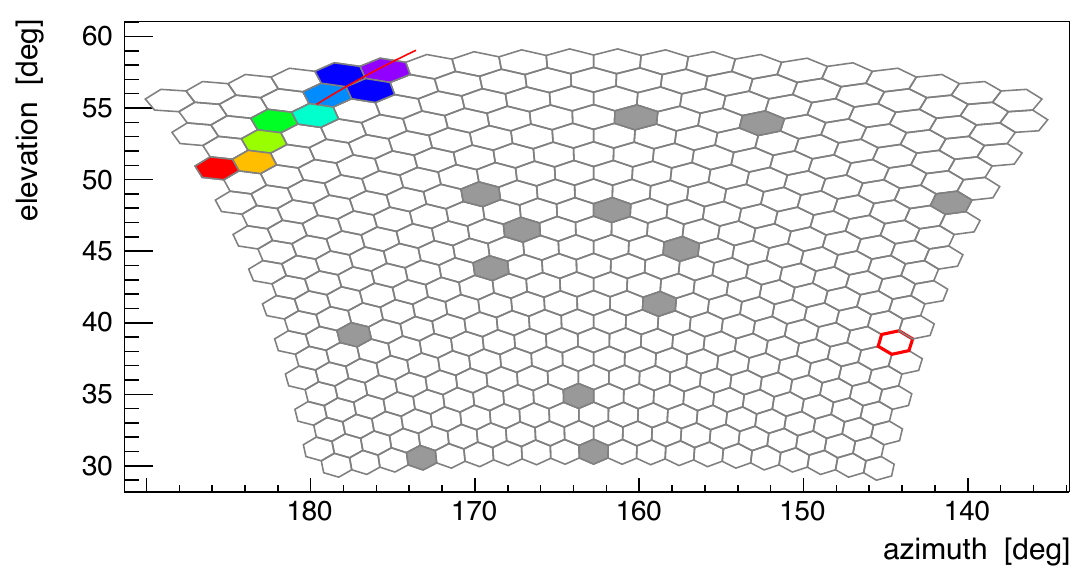}}
\centerline{\includegraphics[scale=0.45]{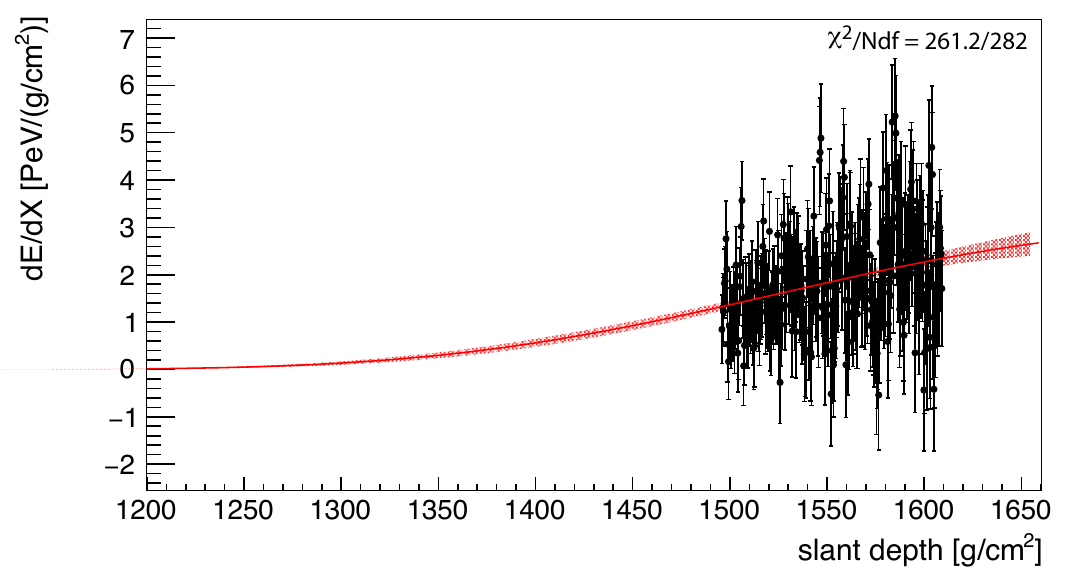}}
\caption{Simulated downward-going UHECR shower that survives all selection and search criteria and has both up and down-ward GF reconstructions, with $L_{down}>L_{up}$. The panels shown and color convention are the same as Fig.\,\ref{fig:backl=1}. 
The reconstructed profile in the bottom panel is that of the rejected upward-going reconstruction.}
\label{fig:simulatedl=0}
\end{figure}
The reconstruction of UHECR-induced air showers with the FD telescopes can also be imprecise in cases in which events trigger only a few pixels of a telescope. This can happen both for low energy showers as well as in cases where the image of the event sweeps across a small portion of the focal plane. In Fig.\,\ref{fig:simulatedl=0} we display an example of such a shower: a simulated iron primary of 6.6 EeV entering the atmosphere at a zenith angle of $82^\circ$ and developing its shower maximum at a depth of 680~g\,cm$^{-2}$.  It is reconstructed with a zenith angle of $\theta_{\rm rec}=85^\circ$ and a shower maximum of $X_{\rm max} = 1206$\,g\,cm$^{-2}$ in the downward mode and mis-reconstructed with $\theta_{\rm rec}=112^\circ$ and $X_{\rm max} = 1755$\,g\,cm$^{-2}$ in the upward mode. For showers in the upward direction we define $X_{\rm max}$ as the slant depth from the exit point of the trajectory to the shower maximum. The GF favors the downward-going geometry so that the discrimination parameter is set to $l=0$. The event is correctly rejected from the signal sample.
This event, with only nine triggered pixels, leaves a similar trace in the camera as the candidate event found in the corner of a HEAT camera ({\it c.f.} Fig.\,2 in the main article).

\section{Reconstruction quality for upward-going showers}

\begin{figure}[t]
\centerline{\includegraphics[scale=0.40]
{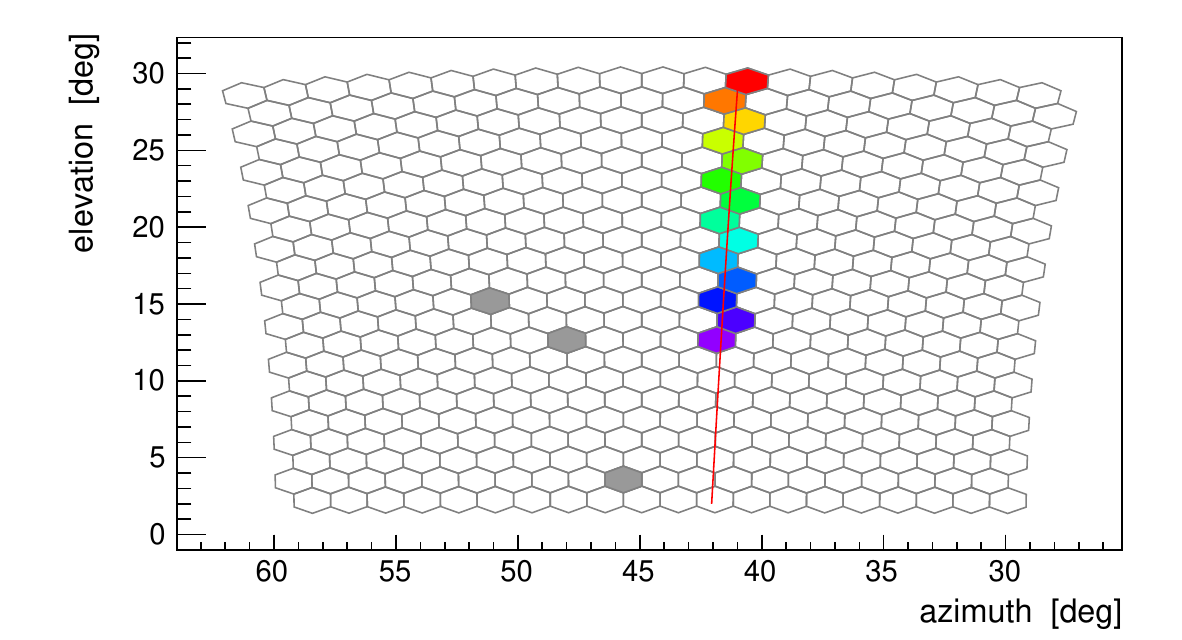}}
\centerline{\includegraphics[scale=0.40]
{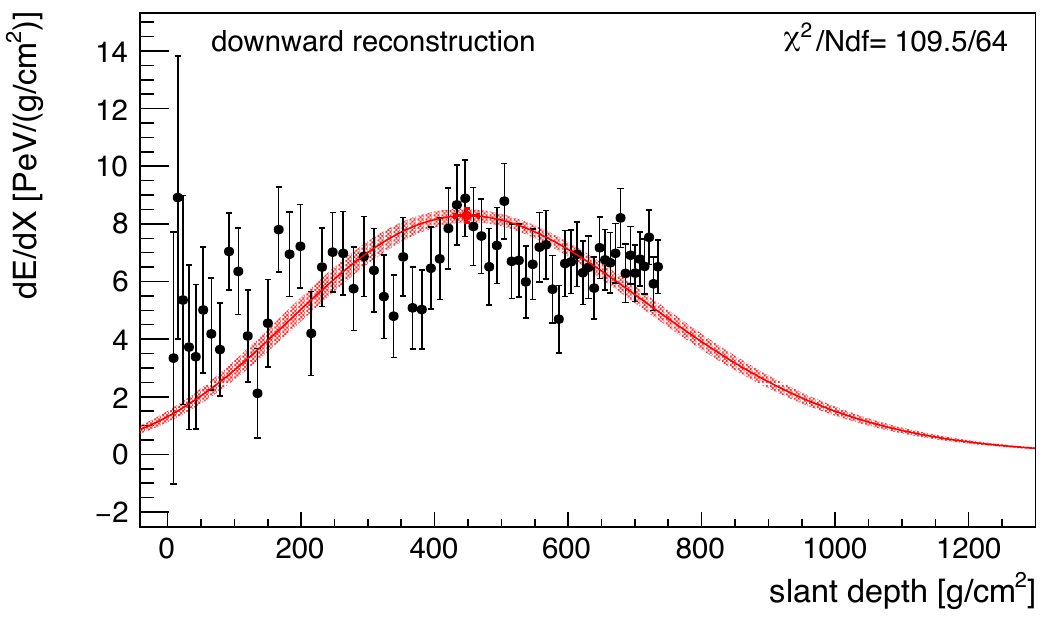}} 
\centerline{\includegraphics[scale=0.40]
{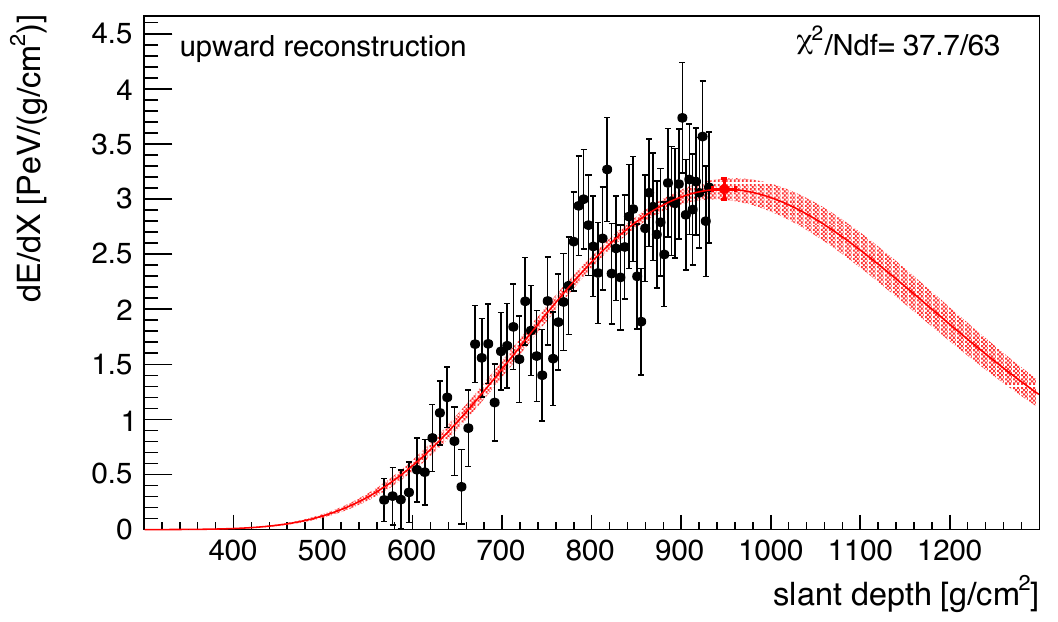}}
\caption{Simulated upward-going shower with ambiguous reconstruction that gives solutions for both upward and downward modes of the Global Fit. The top panel reproduces the information about the triggered pixels with the same color convention as used in Fig.\,\ref{fig:backl=1}. The middle and bottom panels respectively display the best fitted Gaisser-Hillas profiles of the downward and upward modes (see full text). The $\chi^2$ values of each of the two fits are also indicated on the two plots.}
\label{fig:ambiguous}
\end{figure}

An example of a simulated upward-going signal event is shown in Fig.\,\ref{fig:ambiguous}. It has been simulated with a zenith angle of $113^\circ$ and the shower was induced by a 2.4 EeV proton injected 2 km above the ground and it has developed its shower maximum, $X_{\rm max}$, at an atmospheric depth of 1374\,g\,cm$^{-2}$. 
In this case, the GF allows reconstruction both in upward- and downward-going direction and the two reconstructed profiles are compared.
The upward-going reconstruction resulted in a zenith angle of $121.7^\circ$ with a shower maximum at 948 g\,cm$^{-2}$ while the downward-going reconstruction gave $\theta_{\rm rec}=83.6^\circ$ and $X_{\rm max}=448$ g\,cm$^{-2}$. For this event, the upward-reconstruction resulted in a significantly better global fit and the discrimination variable has a value of $l=0.68$. The event is correctly accepted as a signal event.

%



\begin{figure}[t]
\centerline{\includegraphics[scale=0.40]
{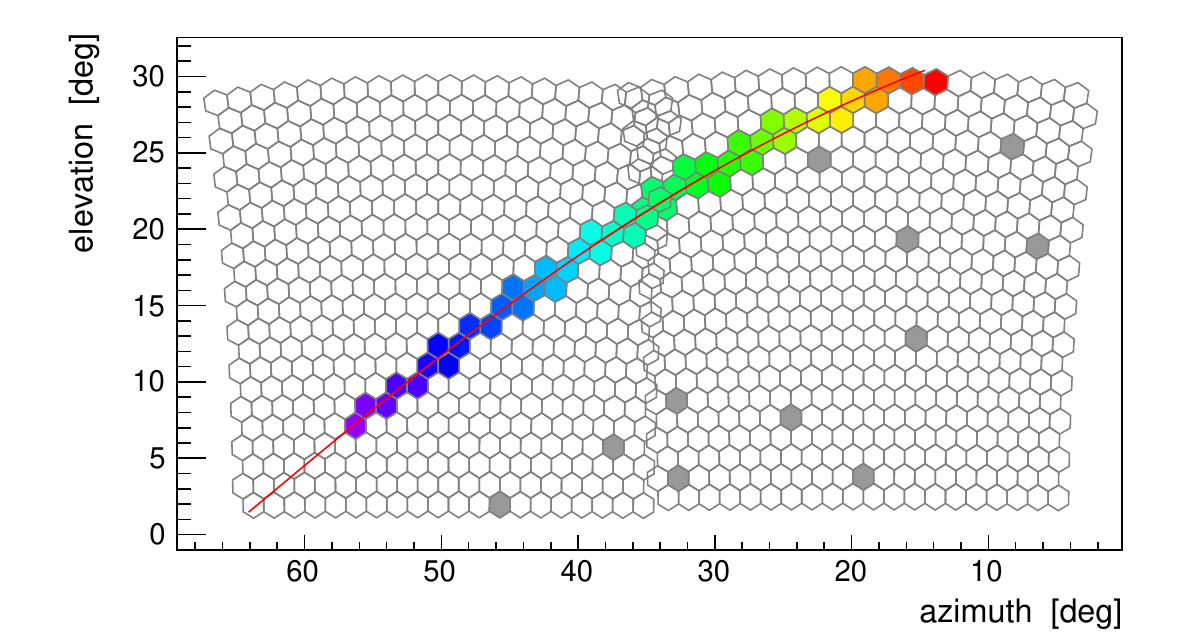}}
\centerline{\includegraphics[scale=0.40]
{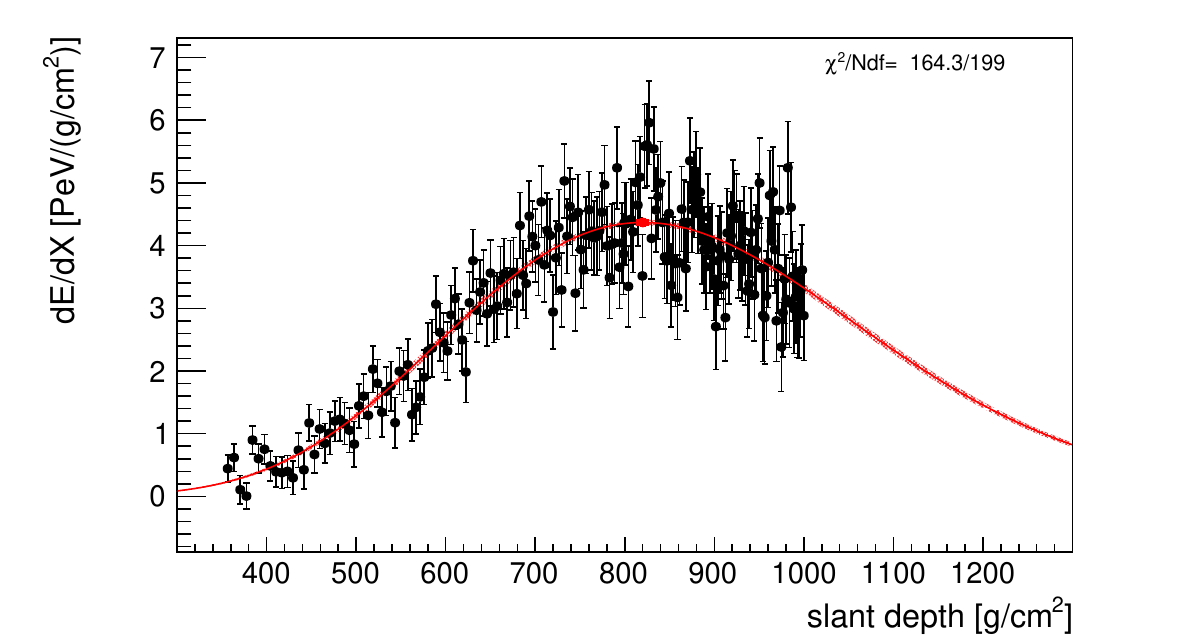}} 
\caption{Simulated upward-going shower that can only be reconstructed in upward-mode. The color code reflects again the start time sequence of the signals (purple first, red last). The bottom panel displays the best fitted Gaisser-Hillas profile (see full text). The $\chi^2$ value of the fit is indicated.}
\label{fig:up-going}
\end{figure}

Another example of a signal event in the zenith angle range of the ANITA events is shown in Fig.\,\ref{fig:up-going}. It was induced by a proton of 3\,EeV going upward with a zenith angle of $\theta_{\rm sim}=114.2^\circ$ and developing its shower maximum at 844\,g\,cm$^{-2}$. 
The camera view shows that the shower track passes through two adjacent FD cameras. The shower can only be reconstructed as upward-going ($l=1$) with a profile well described by a GH shape. Its reconstructed zenith angle and shower maximum are $\theta_{\rm rec}=114.7^\circ$ and 820\,g\,cm$^{-2}$, respectively. Also this event is correctly identified as an upward-going shower.

\begin{figure}[tbh]
\centerline{\includegraphics[scale=0.45]
{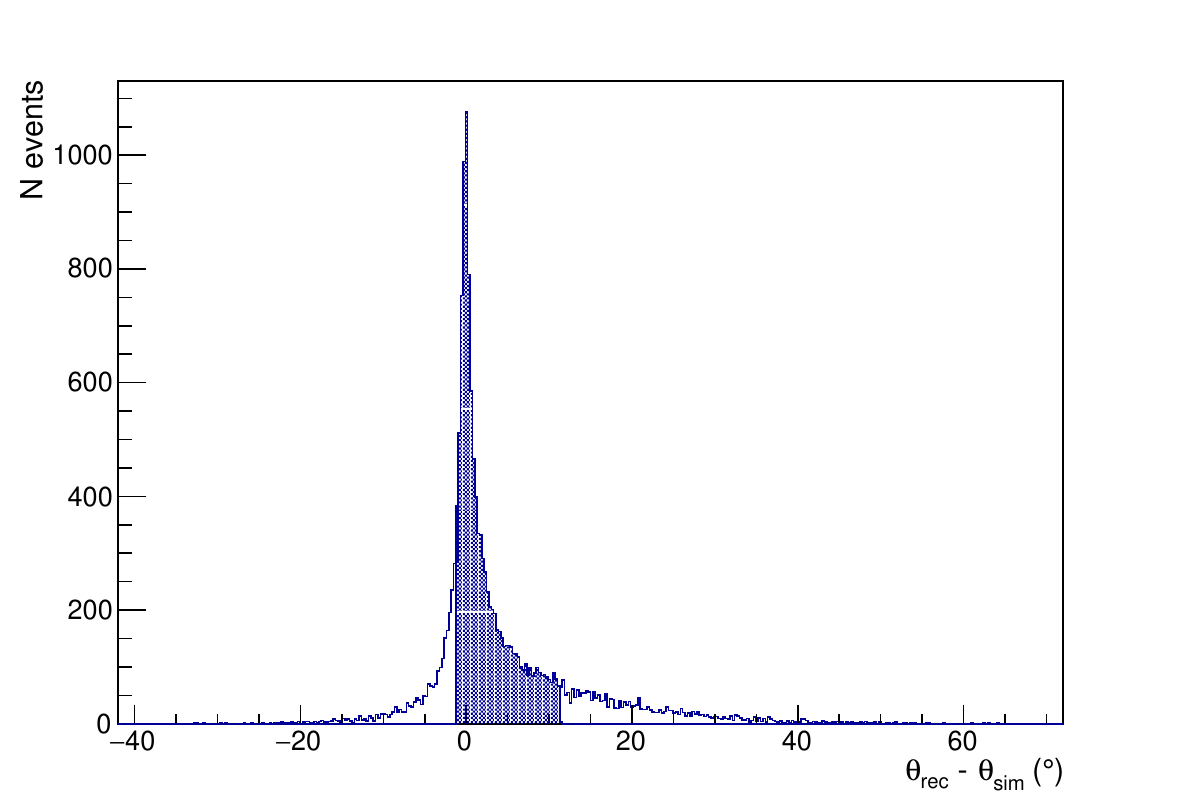}}
\caption{Deviation of reconstructed and simulated zenith angles for upward-going air showers simulated in $\theta_{\rm sim} \in [110^\circ,180^\circ]$.  All selection cuts including $\theta_{\rm rec}>110^\circ$ are applied. The shaded part of the histogram shows the 68\% central interval, covering [$-1.1^\circ, 11^\circ$].}
\label{fig:zenith-resolution-upwards}
\end{figure}

As is demonstrated by these different examples of signal and background events, proper reconstruction of zenith angles for upward-going showers is important to not miss potential signal events. This is studied by signal simulations as part of the exposure calculations. Figure\,\ref{fig:zenith-resolution-upwards} depicts the achieved zenith angle resolution for showers simulated in the zenith angle range $110^\circ \leq \theta_{\rm sim} \leq 180^\circ$, after all cuts are applied, including $\theta_{\rm rec}>110^\circ$. The distribution has an asymmetric shape towards larger reconstructed zenith angles, i.e.\ towards more vertically upward-going showers which remain in the signal sample of upward-going showers.  (We note that the asymmetry to the left tail is biased by the fact that signal simulations only involve showers with $\theta_{\rm sim}>110^\circ$.)
A small fraction of showers reconstructed with $\theta_{\rm rec} < \theta_{\rm sim}$ may be lost by the cut $\theta_{\rm rec}>110^\circ.$ This is accounted for in the calculation of the signal efficiency.

When we consider only the zenith angle range of the anomalous ANITA events, i.e.\ $\theta_{\rm sim} \in [110^\circ,130^\circ]$, the distribution becomes more symmetric and the 68\% central interval reduces to $[-2.96^\circ,9.07^\circ]$. 

\section{Consistency of data and simulations}

To eliminate laser events and UHECR induced showers that are mis-reconstructed as upward-going, a series of cuts is applied as described in the main text. They are tuned to simulated data and - in case of the background sample - verified by a burn sample comprising 10\% of randomly selected data. It is thus important to verify that data and simulations are well understood and agree with one another at the different levels of applied quality cuts.  

\begin{table}[tbh]
\begin{tabular}{|c || c|c|}
\toprule
\text{Successive}	& \text{Fluorescence} & \text{background} \\ 
\text{selection cuts}	& \text{data} & \text{simulation} \\
\hline
\hline
mono pre-selection &  165k     &  279k	\\
GF up-ward mode  &  2774 & 2905 	\\
quality cuts  & 986   &  1157	\\
$\theta >110^\circ$  & 928      & 1064 	\\
$\chi^2_{\rm up}<1.2~\chi^2_{\rm down}$ &  255     &  292	\\
$l>0.55$  &  1     &  0.27	\\
\hline
\end{tabular}
\caption{Events remaining after each upward-going shower selection cut for data compared to the background simulations, the latter weighted with the UHECR energy spectrum.}
\label{tab:EventsAfterCuts}
\end{table}

\begin{figure}[tbh]
\centerline{\includegraphics[scale=0.48]
{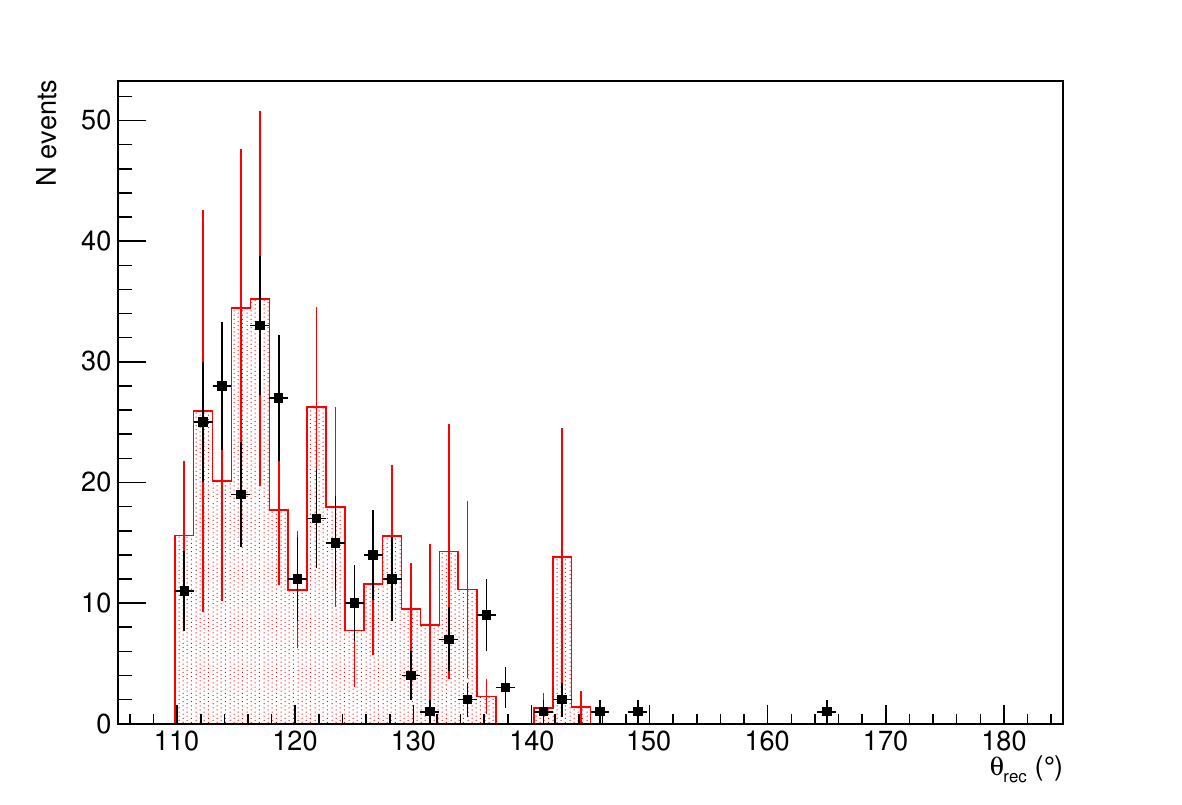}}
\caption{Reconstructed zenith angle distribution both for the data (black symbols) and for the background simulations (red histogram), weighted with the measured UHECR energy spectrum before applying the final cut $l>0.55$.}
\label{fig:zenith-angles}
\centerline{\includegraphics[scale=0.50]
{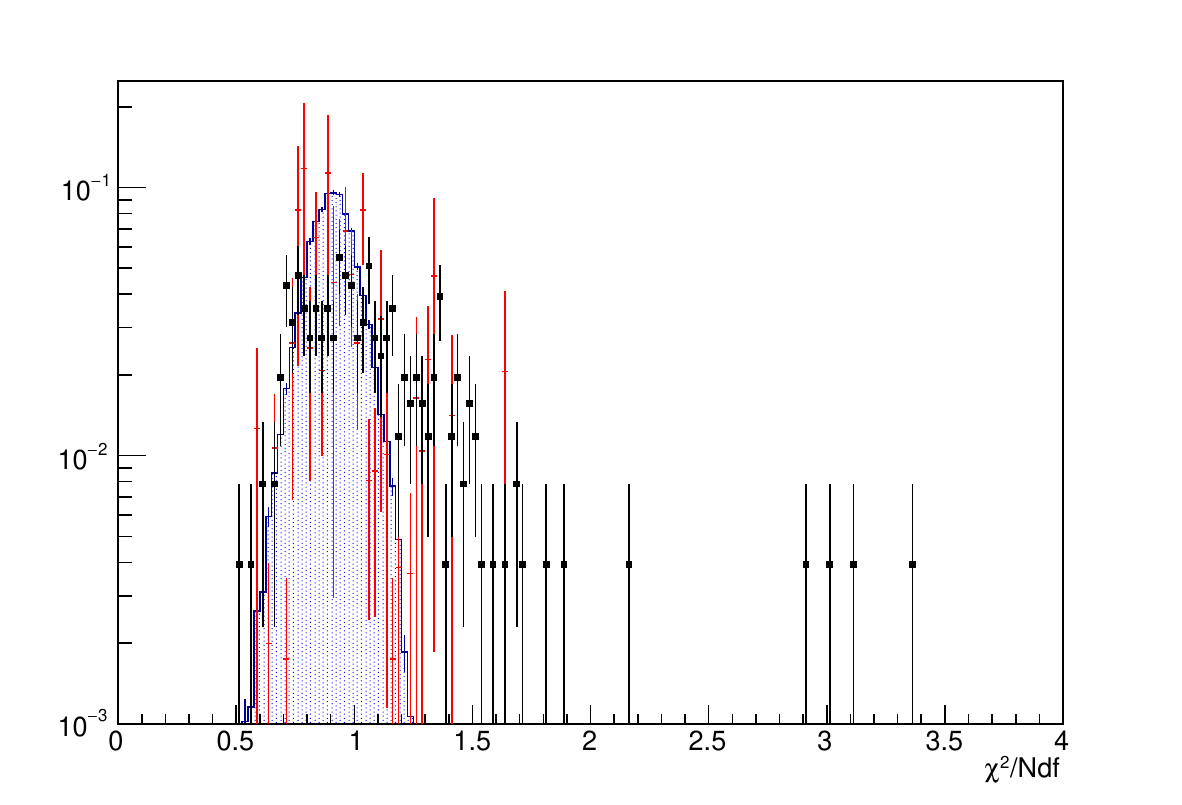}}
\caption{Distribution of $\chi^2$ divided by the number of degrees of freedom, Ndf, from the GF applied to data (black symbols), simulated background (red markers), and simulated upward-going events (blue histogram), all after all selection cuts, just before applying a cut on the discrimination parameter $l$. Simulated background events were weighted with the measured UHECR spectrum, while simulated signal events were weighted with a $E^{-1}$ power-law distribution. For better comparison of shapes, all histograms are normalized to one.}
\label{fig:chi2-plot}
\end{figure}

The effect of progressively applying the cuts as described in the main article is presented in Table~\ref{tab:EventsAfterCuts} both for data and simulations. The agreement in the succession of cuts after the GF has been applied is within known uncertainties. 
The limited number of simulated background events and their corresponding weighting to the UHECR spectrum are subject to uncertainties connected to the weighting and are at the level of $\sim 10 \%$. Also, the light collection efficiency (calibration) of the FD telescopes response modeled in the MC simulations is quoted to have an uncertainty of $9.9\%$~\cite{Dawson:2020bkp}. This   
accounts for a good part of the differences after the cuts are applied beyond the first level. 
At the first level shown, the total number of passing events is dominated by faint, predominantly low energy events and the exposure to them is very sensitive to small discrepancies between simulations and reality. 

As another verification of data and background simulations, we present in Fig.\,\ref{fig:zenith-angles} the distribution of reconstructed zenith angle for data (black points) and simulated background (red histogram), the latter weighted with the measured UHECR spectrum, before the final cut $l>0.55$ is applied. Again, the agreement between the distributions is found to be very good. 

Finally, we inspect in Fig.\,\ref{fig:chi2-plot} the $\chi^2$-distributions obtained from the GF, now for data and simulated background, in comparison to that from the simulated signal events, again before the cut on $l$ is applied. As before, the simulated background events are weighted with the UHECR spectrum, while the simulated signal events are weighted according to a $E^{-1}$ power law distribution.
These distributions illustrate once more good agreement between data and background simulations. Since the background is by construction a selection of mis-reconstructed events, we expect the data and background distributions to exhibit on average larger $\chi^2$-values than the signal simulations. This is what is found:
\begin{table}[h!]
    \centering
    \begin{tabular}{lll}
          & mean & sigma \\ \hline
data      & $1.07\pm0.02$  & $0.37\pm0.01$\\
background& $0.94\pm 0.02$ & $0.19\pm 0.2$\\
signal    & $0.9067\pm 0.0005$ & $0.1145\pm0.0005$
    \end{tabular}
    \caption{First and second moment of the $\chi^2$-distributions from the GF applied to data, simulated background, and simulated signals.}
    \label{tab:chi2}
\end{table}

The difference between data and background simulation is also not unexpected and is mostly caused by non perfect detector simulations. 


\end{document}